David Lopez Del Moral Hernandez; Andres Barrado Bautista; Clara Marina Sanz Garcia; Antonio Lazaro Blanco; Pablo Zumel Vaquero (2020).

Analysis, design, and implementation of the AFZ converter applied to photovoltaic systems

# Analysis, design and implementation of the AFZ converter applied to photovoltaic systems

David López del Moral, Andrés Barrado, Marina Sanz, Antonio Lázaro, Pablo Zumel
Power Electronics System Group, University Carlos III of Madrid,
e-mail: andres.barrado@uc3m.es

***Abstract*- Grid-tied photovoltaic (PV) installations with Distributed Maximum Power Point Tracking (DMPPT) architectures include a DC-DC Module Integrated Converter (MIC) for managing each PV panel, isolating it from the others, reducing the mismatching effect and maximizing the harvested power. In this paper, the Autotransformer Forward converter with type-Zeta resonant reset (AFZ) is proposed as a DMPPT architecture's MIC candidate. The main characteristics of the AFZ converter are the high versatility due to its voltage step-up and step-down capability; the use of an optimized autotransformer with only two windings, reducing the complexity and power losses of this component; the good dynamic performances, like the Forward converter ones; the low number of components and the simplicity and high feasibility associated to the use of just one active switch. Besides, soft switching transitions are achieved thanks to the autotransformer type-Zeta resonant reset. The steady-state theoretical analysis, considering the effect of the autotransformer leakage inductance, is presented. The converter is also studied in the frequency domain, obtaining the small-signal transfer functions. A design procedure based on the requirements of a 100 kW grid-tied photovoltaic installation is described, yielding in a 225 W prototype with efficiencies up to 95.6 %. Experimental results validate the theoretical analysis.**

## I. INTRODUCTION

Nowadays, the interest in green and renewable energy sources is increasing. One of the most relevant is the Photovoltaic energy. As shown in Figure 1, the world PV Cell/Module production presents an exponential growth from 2010 to 2018 [1]. In Figure 2, the annual PV installations in the EU and candidate countries exhibit a peak in the years 2010-2012 and sustained value from 2014. This evolution shows that PV is becoming a significant part of the energy mix in many countries. These facts, in conjunction with social awareness, make this field worthy and motivate research in the improvement of current PV conversion technologies and the development of new ones.

In high power grid-tied PV installations, one of the significant issues comes from the reduction in the harvesting power due to differences between the PV panels connected to the same string. This issue is commonly known as mismatching.
As long as the PV panels are directly connected to the string, any difference in the electrical characteristic of one of them influences the others, reducing the power generated by them, even with full irradiance and optimum environmental conditions.

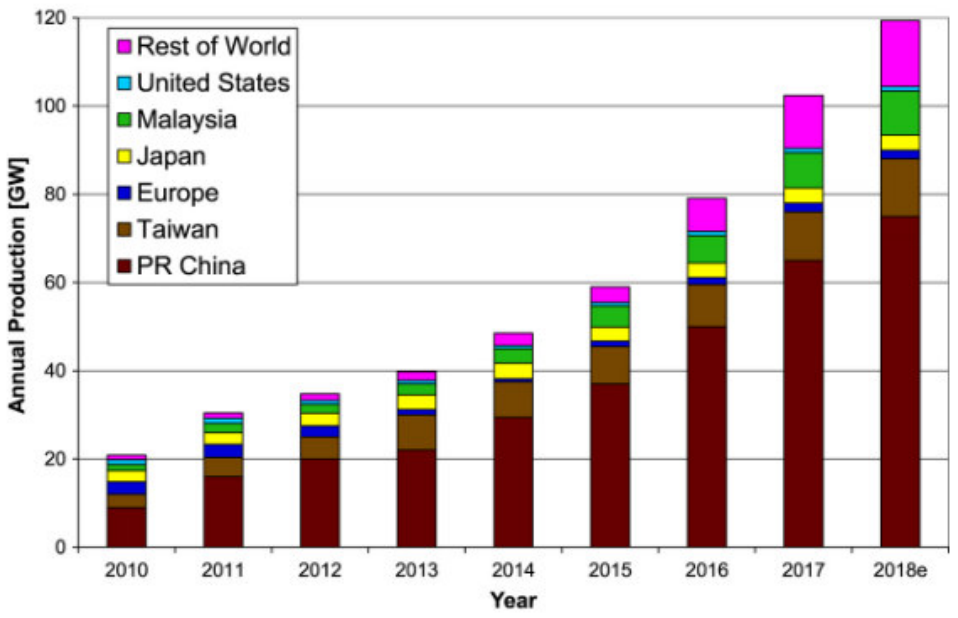

Figure 1. World PV Cell/Module Production from 2010 to 2018 [1]

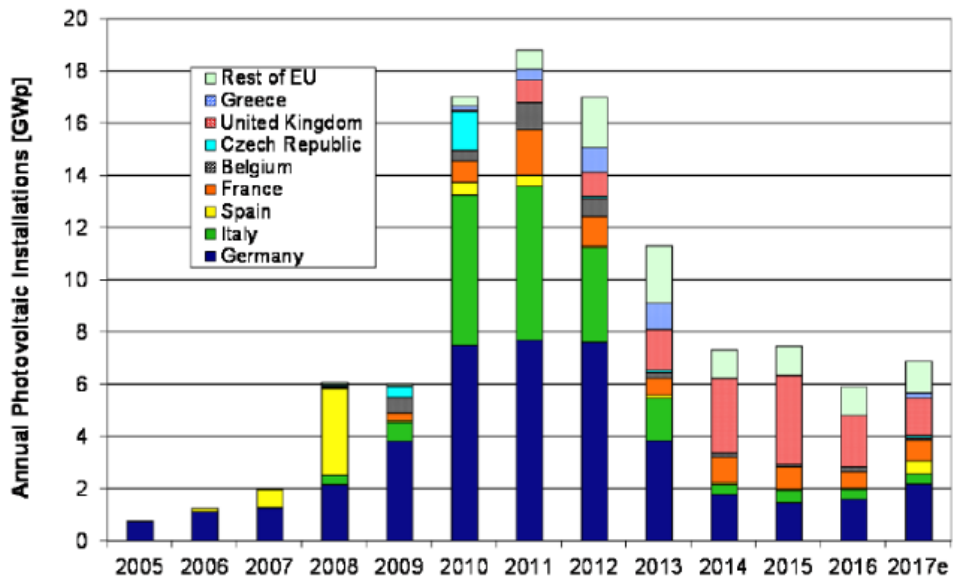

Figure 2. Annual PV installations in EU and candidate countries [2]

There are many causes that modify the electrical characteristics of a PV panel, such as dirt, shadows, aging, temperature differences, etc., [2] - [3]. How all these causes affect the PV panel electrical characteristics is an interesting research topic, as shown in [4] - [7].

One of the most popular solutions to overcome the mismatching issue is based on attaching a DC-DC module integrated converter (MIC) to each PV panel. This configuration, especially when step-up and step-down MICs are used, allows generating the maximum available power regardless of the conditions of the rest of the PV panels in the installation [28] – [32]. This MIC also performs the control of the PV panel and operating it on its maximum power point (MPP).

The architectures that implement this solution are denoted as Distributed Maximum Power Point Tracking (DMPPT) architectures, [8] - [14], see Figure 3. As a drawback, DMPPT architectures require a high number of MICs, one per PV panel, increasing the cost of the installation. Therefore, for making the PV installation flexible and profitable, the MIC requirements are low cost, high efficiency, and the voltage step-up and step-down capability.

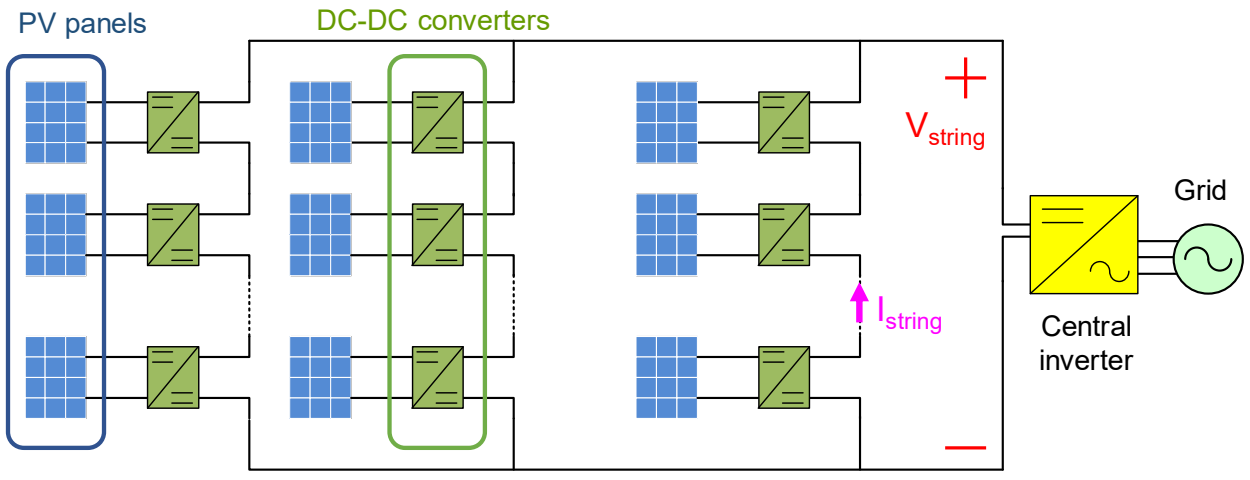

Figure 3. PV grid-tied installation with the DMPPT architecture

Several authors have focused their research on improving the efficiency of the MIC. For this purpose, one of the most interesting approaches focuses on the reduction of the power losses, not processing all the power delivered by the MIC to the load. This principle can be found in the literature as Partial-Power Conversion (PPC), Series Connection, Parallel-Power-Processed (PPP), or Direct Energy Transfer (DET) converters [15] – [23]. In all of them, the efficiency improvement is achieved due to the converter only manages a part of the energy, whereas the rest of the output power is directly delivered from the PV panel to the load. The best efficiency achieved in this type of converters is up to 98 %, as shown in [17]. The main limitation of this type of topologies is that they are only capable of voltage step-up. This limitation reduces the flexibility when designing a DMPPT architecture. Beyond the topologies that do not process the full power, other authors have also obtained efficiencies around 98 % with full power processing topologies, but also with the limitation of not being capable of both voltage step-down and step-up [25] - [27].

It has been demonstrated that the highest flexibility regarding the number of PV panels per string is only achieved with voltage step-up and step-down converters [28] - [32]. Within the voltage step-up and step-down topologies, one of the most promising ones is the classical Non-Inverting Buck-Boost converter (also known as the Four-Switch Buck-Boost converter), see [28] and [32], in which very high efficiencies have been obtained. However, this topology has some drawbacks such as high current through the inductor and switches, and the need for four switches and drivers. Therefore, the complexity and components count increases.

The Autotransformer Forward converter with type-Zeta resonant reset (AFZ) converter, introduced in [33] and deeply analyzed in this paper, overcomes the mismatching issue, with the voltage step-up and step-down capability advantage. Compared to other solutions from the literature, this converter is simpler, with only one active switch and driver. On the other hand, the efficiency of this converter fits with those found in the literature, although slightly lower than the one described in reference [28] and [48]. Another interesting characteristic of the AFZ converter is the type-Zeta resonant reset that yields in the autotransformer optimization, the possibility of avoiding additional snubber networks, and the obtaining soft-switching characteristics. Besides, this topology does not present right-half-plane (RHP) zeros in its main small-signal transfer functions, which allows obtaining better dynamic performances. The AFZ converter was previously introduced in [33], but including only the steady-state analysis of the converter's ideal model. In this work, a more accurate steady-state analysis is included, with a more realistic converter model that considers the effects of the autotransformer leakage inductance. This consideration reveals an additional switching interval. Moreover, the frequency domain analysis is carried out, and the main small-signal transfer functions are obtained.

The paper is structured as follows. The principle of operation and the time domain and frequency domain theoretical analysis are carried out in Section II. Section III details the design procedure for the AFZ converter, being applied to a 100 kW grid-tied PV installation, which takes into account the effect of the mismatching. The experimental results of a 225 W prototype are also shown in this Section for verifying the theoretical analysis. Finally, the conclusions of this research are summarized in Section IV.

## II. THEORETICAL ANALYSIS OF THE AFZ CONVERTER

The AFZ converter can be seen as a Forward based converter but replacing the Forward's transformer for a two windings autotransformer, see Figure 4. In comparison to an equivalent transformer, the size and power losses of the autotransformer are considerably lower, due to only a part of the output power is magnetically processed, as is described in Section II.C. The connection of the autotransformer allows a percentage of the delivered power not to be magnetically processed. On the other hand, in this converter, isolation is lost. This fact is not considered as a drawback, due to it is not usually required in MIC for PV applications.

As it is highlighted in cyan color in Figure 4, the autotransformer reset is carried out employing a type-Zeta resonant reset network. This network is a key point in the AFZ converter as it avoids the need to have a third winding in the autotransformer to reset it. In addition, this network allows its core to be excited symmetrically between the first and third quadrants of the B-H plane, reducing the autotransformer size. Moreover, it avoids the use of snubber networks.

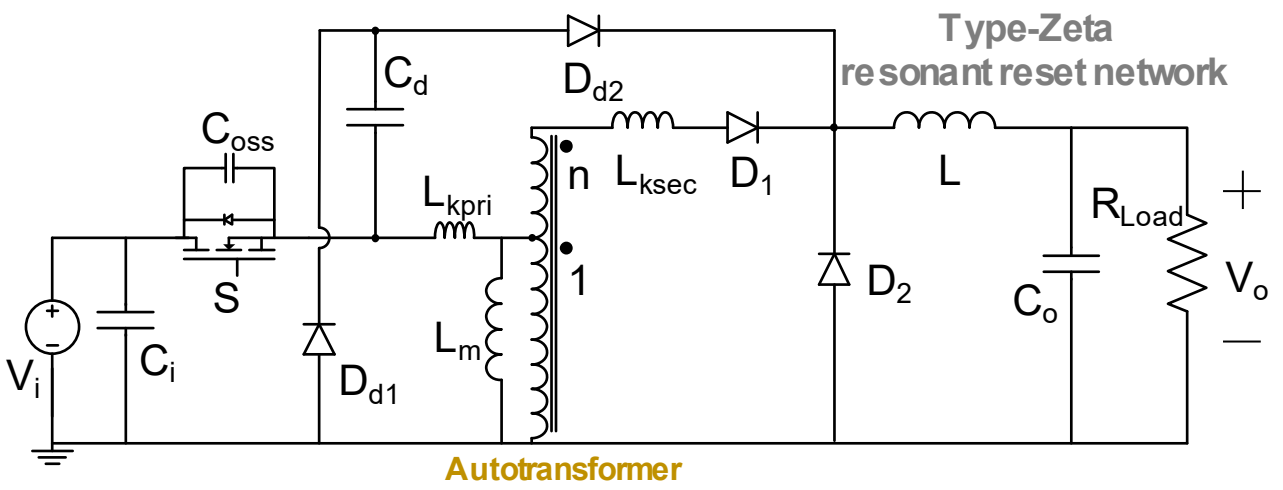

Figure 4. AFZ converter electrical scheme

Although the AFZ converter operation principle and its main steady-state waveforms were introduced in [33], in that publication, the ideal model was considered. This paper studies a more realistic model, considering the leakage inductances of the autotransformer $L_{kpri}$ and $L_{ksec}$, as well as the equivalent drain-source MOSFET parasitic capacitance $C_{oss}$. The steady-state analysis includes not only more accurate waveforms but also the main expressions that define the voltage and currents of the AFZ converter components. Regarding the principle of operation, an additional switching interval related to the leakage inductances is defined. Besides the aforementioned additional information, in comparison to [33], the small-signal model of the AFZ converter is carried out, and the main small-signal transfer functions are obtained in continuous conduction mode (CCM), which are needed when designing the converter compensator. Also, more experimental results are included in this paper.

### *A. Principle of operation*

The switching period is divided into two main intervals, depending on the MOSFET state, the $t_{ON}$ and $t_{OFF}$ intervals. The $t_{OFF}$ interval is further divided into three subintervals, depending on the magnetizing inductance ($L_m$) current and the resonant capacitor ($C_d$) voltage state. Finally, the transitions between the two main intervals are also analyzed. Every interval, subinterval, and transition interval have a figure associated, highlighting the paths followed by the current. These paths are depicted with thick blue lines, and arrows denote the direction of the current. The rest of the converter components are depicted with thinner grey lines.

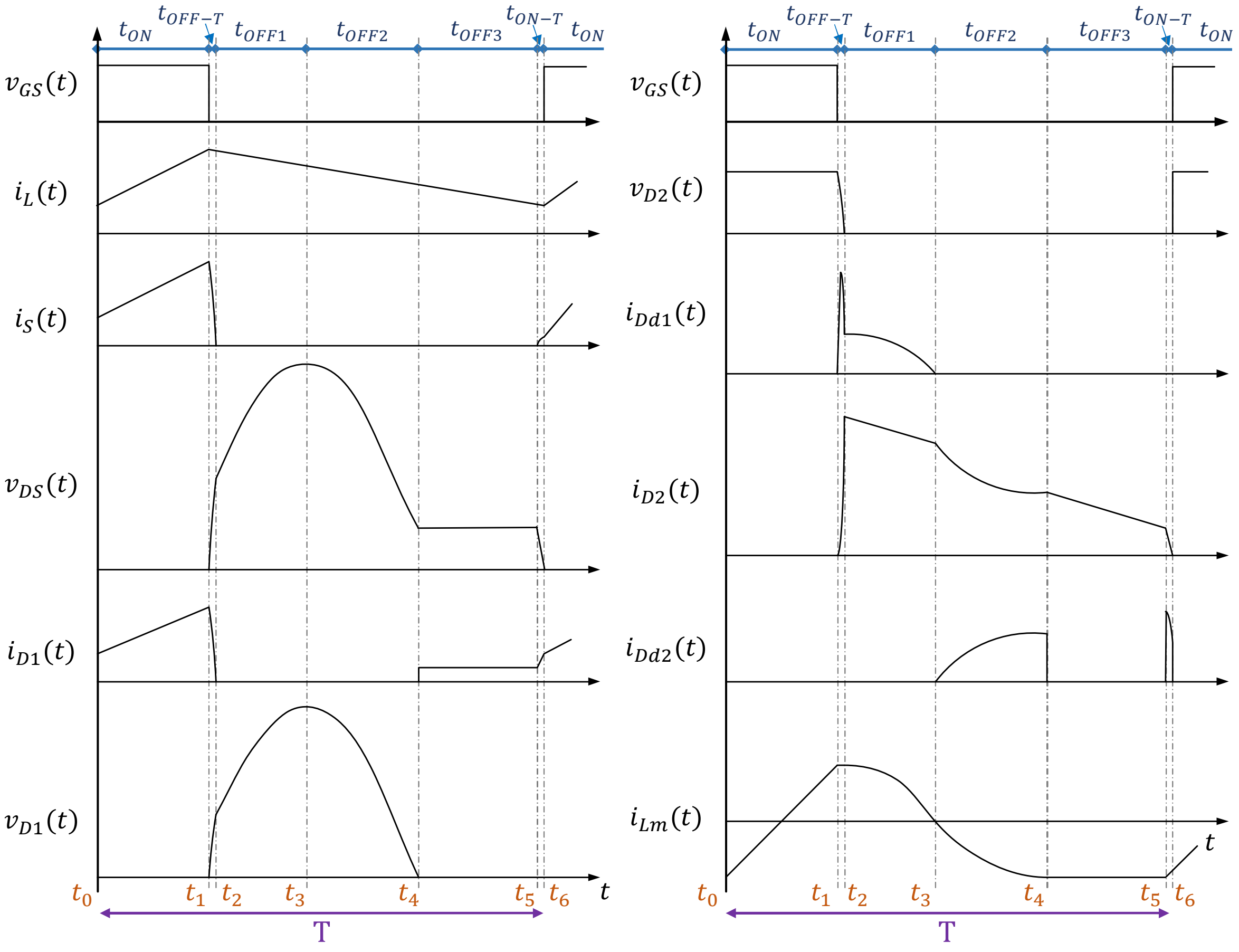

Figure 5. Main waveforms of the AFZ converter in continuous conduction mode (CCM)

The main voltage and current waveforms of the AFZ converter are depicted in Figure 5 for a better understanding of the principle of operation. Note that the time intervals and the voltage and current values are not to scale. Considering this, the $V_{oss}(t)$ voltage waveform is exactly the same as $V_{DS}(t)$. The $V_{cd}(t)$ waveform is similar to $V_{DS}(t)$, but with a negative offset, as it is shown in Section II.B.

- $t_{ON}$ ($t_0$ - $t_1$)

The first of the intervals is denoted as $t_{ON}$ and occurs while the S MOSFET is turned-on, see Figure 6. During this time, the converter delivers the power from the input to the output through the autotransformer. The use of an autotransformer gives the current a direct path between the input voltage source and the output load, resulting in a not-magnetically processed power transference. For further detail regarding the power processing in the autotransformer, see Section II.C. Due to the current flow, the AFZ converter inductances L, $L_m$, $L_{kpri}$, and $L_{ksec}$ increase their energy. This interval ends when the S MOSFET is turned-off, leading to the turn-off transition, $t_{OFF\text{-}T}$, interval.

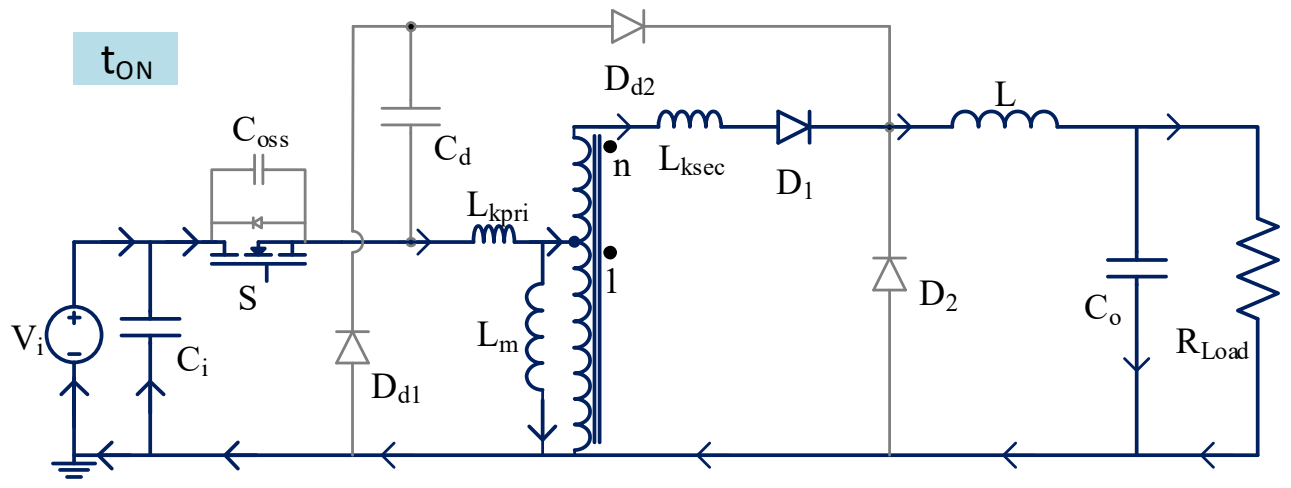

Figure 6. Current paths in the AFZ converter during the $t_{ON}$ interval

- $t_{OFF\text{-}T}$ ($t_1 – t_2$)

This interval lasts a very short time, during the reset of the $L_{kpri}$ and $L_{ksec}$ leakage inductances. The previously stored energy in the leakage inductances is abruptly delivered to the $C_{oss}$ and $C_d$ capacitors, see Figure 7, resulting in a high-frequency resonance. The energy transferred to the capacitors increases their voltages, generating a voltage step. The amplitude and slope of these voltage steps are limited by $C_d$ and $C_{oss}$ capacitances.

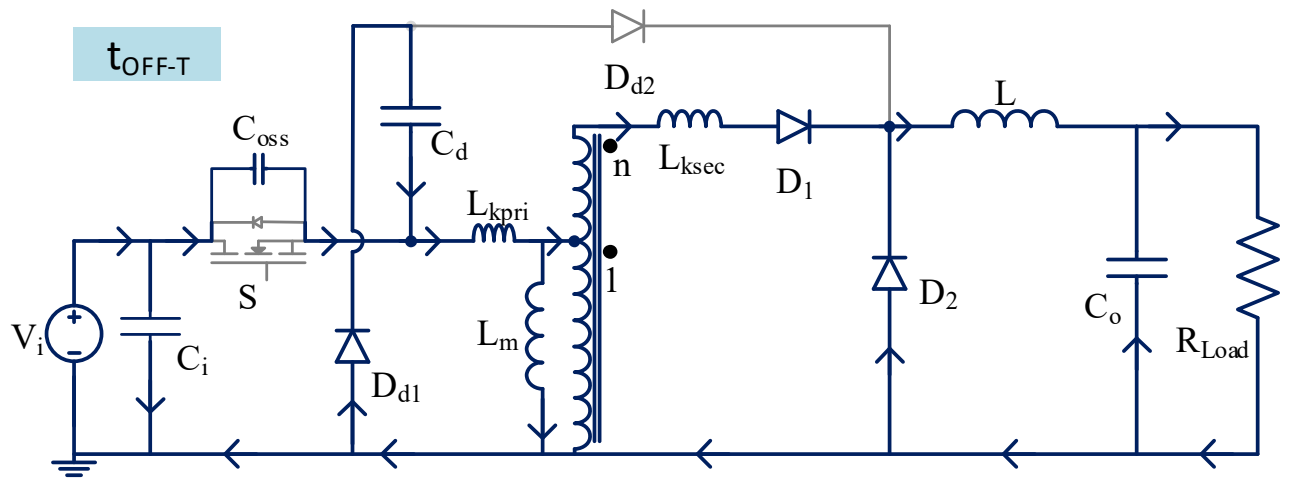

Figure 7. Current paths in the AFZ converter during the $t_{OFF\text{-}T}$ interval

During this interval, the magnetizing inductance $L_m$ is also delivering its energy to the capacitors mentioned above, although it occurs in a much lower frequency, and this effect can be therefore neglected.

- $t_{OFF1}$ ($t_2 – t_3$)

Once the $L_{ksec}$ leakage inductance delivers its energy to the $C_{oss}$ and $C_d$ capacitors and $L_{kpri}$ current equals the magnetizing current, the $t_{OFF1}$ subinterval begins. During this subinterval, the magnetizing and the $L_{kpri}$ inductances continue delivering their energy to the capacitors by a resonant process, see Figure 5. It is noteworthy that the energy delivered by the $L_{kpri}$ inductance is almost negligible in comparison to the magnetizing inductance energy. This interval is considered as the first part of the Type-Zeta resonant reset. Although four components take place in the resonant reset process, the elements that fundamentally affect resonance are $L_m$ and $C_d$, due to they are orders of magnitude higher than $L_{kpri}$ and $C_{oss}$.

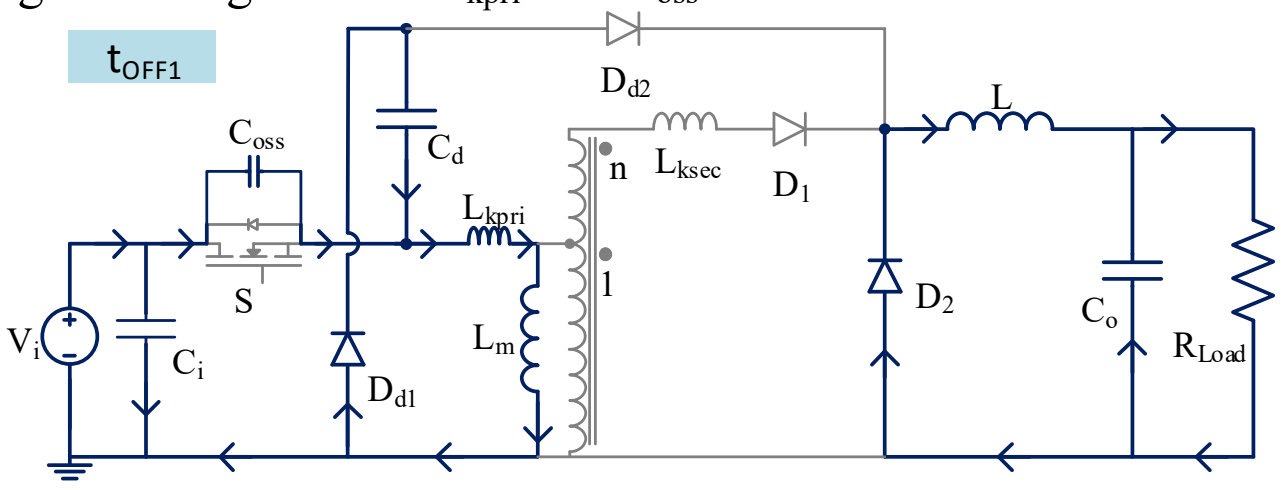

Figure 8. Current paths in the AFZ converter during the $t_{OFF1}$ interval

The S MOSFET and $D_1$ diode reach their maximum voltages at the end of this subinterval when the magnetizing current gets zero amperes. During this subinterval, Figure 8, the output inductance is also delivering its previously stored energy to the load through the $D_2$ diode.

- $t_{OFF2}$ ($t_3$ - $t_4$)

The second part of the Type-Zeta resonant reset takes place during the $t_{OFF2}$ subinterval. As can be seen in Figure 9, the capacitors are delivering the stored energy back to the magnetizing inductance $L_m$. It is noteworthy that the $I_{Lm}$ current direction changes its polarity. The charge that $C_d$ and $C_{oss}$ capacitors give back to $L_m$ flows through the $D_{d2}$ diode and the output filter. In Figure 5 can be seen that the $I_{Dd2}$ current contribution does not disturb the $I_L$ current waveform. The one affected is the $I_{D2}$ current waveform.

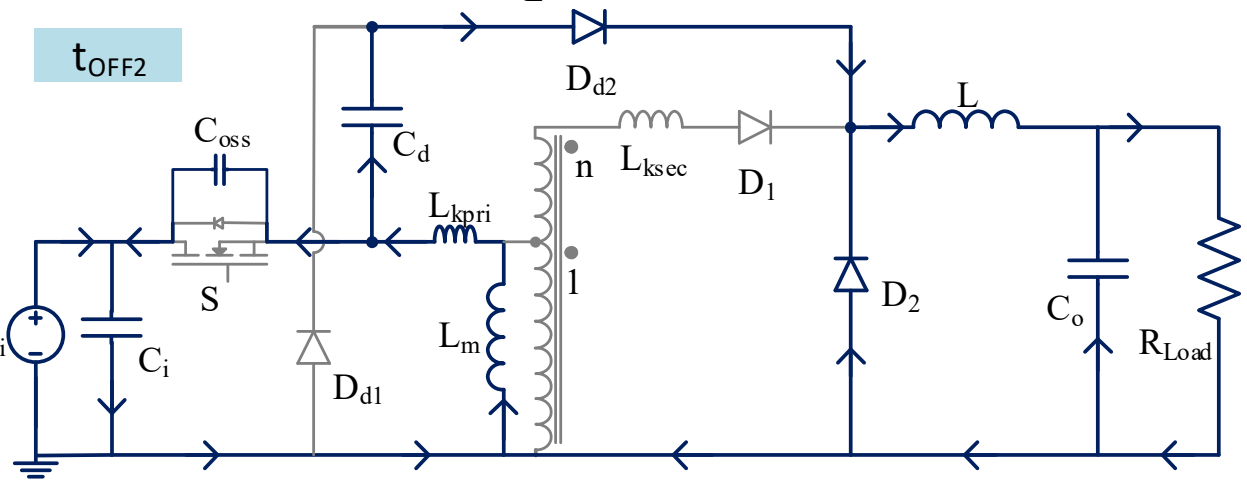

Figure 9. Current paths in the AFZ converter during the $t_{OFF2}$ interval

- $t_{OFF3}$ ($t_4$ - $t_5$)

Once the Type-Zeta resonant reset is finished, the $t_{OFF3}$ subinterval takes place. This subinterval remains from the moment when the $C_d$ and $C_{oss}$ capacitors deliver all their previously stored energy back to the magnetizing inductor until the time when the S MOSFET is turned on again, leading to the $t_{ON-T}$ transition interval. During this interval, the voltage at the autotransformer primary side becomes zero volts, and the current through the magnetizing inductance remains constant with negative polarity. As can be seen in Figure 10, the magnetizing current flows through the primary and secondary windings. Therefore, a part of the magnetizing inductance current, defined by $\frac{1}{1+n}$, flows through the $D_1$ diode. This fact leads to Zero Voltage Switching characteristics on this diode when S MOSFET is switched on.

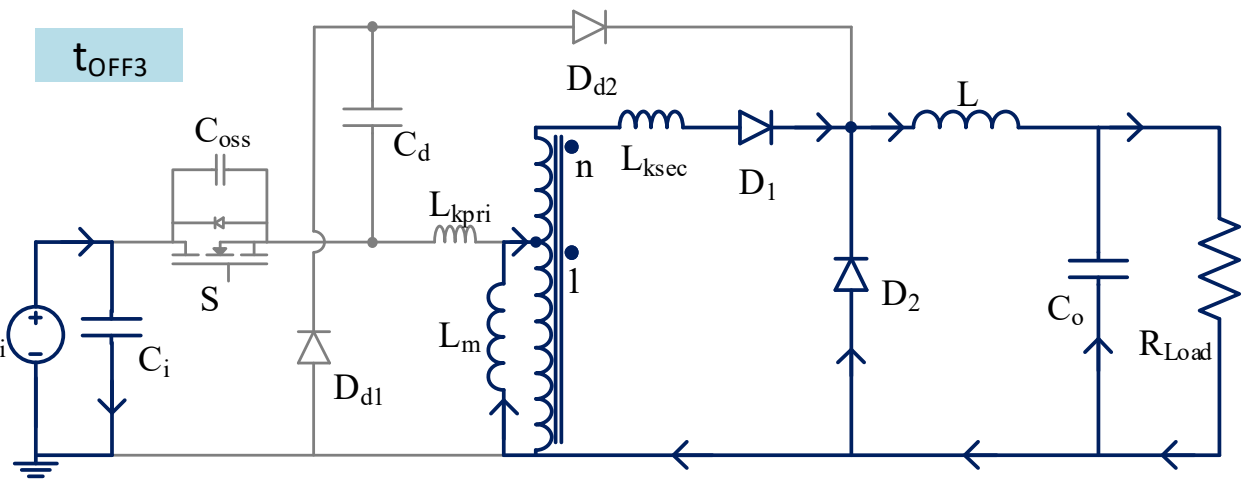

Figure 10. Current paths in the AFZ converter during the $t_{OFF3}$ interval

The expression (1) defines the maximum duty cycle $D_{max}$, which corresponds with the minimum $t_{OFF3}$ duration, as a function of the reset resonant frequency $f_{res}$ and the switching frequency $f_{sw}$.

$$D_{max} = \frac{T_{sw} - 0.5 \cdot T_{res}}{T_{sw}} = \frac{2 \cdot f_{res} - f_{sw}}{2 \cdot f_{res}} \quad (1)$$

Being

$$f_{res} \cong \frac{1}{2 \cdot \pi \cdot \sqrt{(L_m + L_{kpri}) \cdot (C_d + C_{oss})}} \cong \frac{1}{2 \cdot \pi \cdot \sqrt{L_m \cdot C_d}} \quad (2)$$

For design purposes, the parasitic capacitance $C_{oss}$ and the leakage inductance $L_{kpri}$ are neglected due to they are in practice orders of magnitude lower than the resonant capacitance $C_d$ and the magnetizing inductance $L_m$.

- $t_{ON-T}$ ($t_5$ - $t_6$)

When the S MOSFET is turned-on again, the transition interval between the $t_{OFF}$ and $t_{ON}$ intervals takes place; this transition is denoted as $t_{ON-T}$. During this transition, the current flows through the S MOSFET. However, due to the leakage inductances $L_{kpri}$ and $L_{ksec}$ do not allow the current to change abruptly, a part of this current follows the path formed by the $C_d$ capacitor, the $D_{d2}$ diode, and the output filter, see Figure 11. This current flow, which charges the $C_d$ capacitor with negative polarity, remains until the current through the $D_1$ equals the output inductor current. At the end of the $t_{ON\text{-}T}$ interval, the current through the $D_2$ diode becomes zero. It is noteworthy that the capacitor $C_d$ is charged with negative polarity just before the next switching cycle begins.

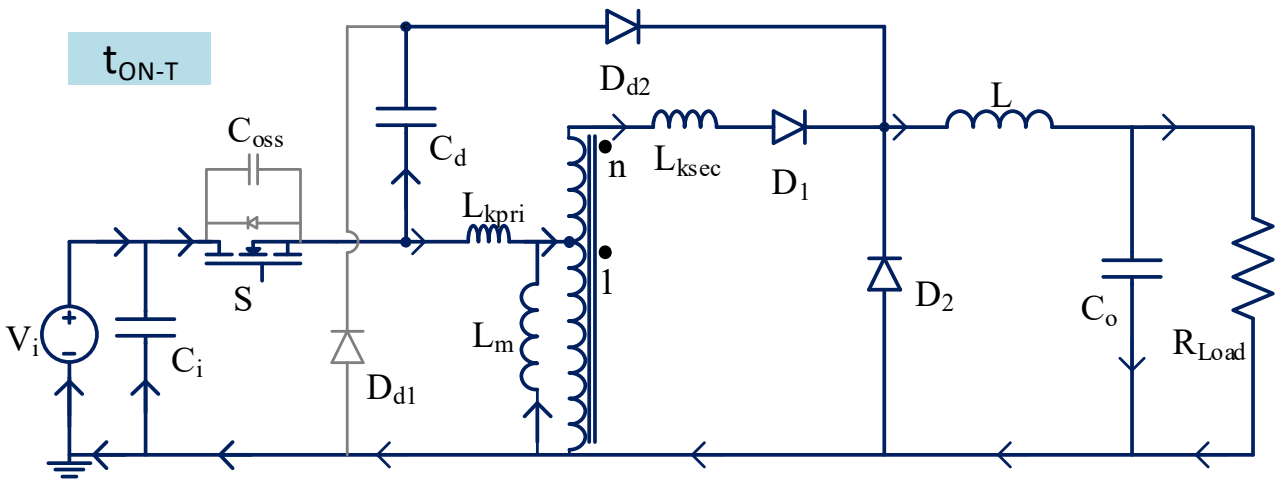

Figure 11. Current paths in the AFZ converter during the $t_{ON\text{-}T}$ interval

Table I summarizes the main events in each switching interval.

## B. *Steady-state operation in continuous conduction mode*

In this section, the most relevant voltages and currents expressions for the main components of the AFZ converter are detailed, as well as their waveforms. Previously, the AFZ converter output-input voltage transfer function is introduced in expression (3), obtained by applying the voltage-per-second balance to the output filter inductor L.

$$\frac{V_o}{V_i} = (1 + n) \cdot D \tag{3}$$

As can be seen, the output-input voltage transfer function is like the Forward converter one, but including the effect of the autotransformer connection, see the $(1 + n)$ factor.

Table I. Summary of the principle of operation for the AFZ converter

| Switching interval | | Start event | Main considerations | Final event |
|---|---|---|---|---|
| $t_{ON}$<br>$(t_0$ - $t_1)$ | | S is turned-on | L, $L_{kpri}$, $L_{ksec}$ and $L_m$ → store energy<br>Part of the output power → not magnetically processed<br>$D_1$ → positive biased | S is turned-off |
| $t_{OFF\text{-}T}$<br>$(t_1$ - $t_2)$ | | S is turned-off | Very short-time interval<br>$L_{kpri}$ and $L_{ksec}$ → deliver energy to $C_d$ and $C_{oss}$<br>$D_{d1}$, $D_1$ and $D_2$ → positive biased | $i_{Lkpri} = I_{Lm}$<br>$i_{Lksec} = 0$ A |
| $t_{OFF}$<br>$(t_2 – t_5)$ | $t_{OFF1}$<br>$(t_2 – t_3)$ | $i_{Lkpri} = I_{Lm}$<br>$i_{Lksec} = 0$ A | The first part of the Type-Zeta resonant reset<br>L, $L_{kpri}$ and $L_m$ → deliver energy<br>$D_{d1}$ and $D_2$ → positive biased | $i_{Lkpri} = 0$ A<br>$i_{Lm} = 0$ A |
| | $t_{OFF2}$<br>$(t_3 – t_4)$ | $i_{Lkpri} = 0$ A<br>$i_{Lm} = 0$ A | The second part of the Type-Zeta resonant reset<br>$L_{kpri}$ and $L_m$ store energy from $C_d$ and $C_{oss}$<br>L → deliver energy<br>$D_{d2}$ and $D_2$ → positive biased | $v_{Cd} = 0$ V<br>$v_{Coss} = V_i$ |
| | $t_{OFF3}$<br>$(t_4 – t_5)$ | $v_{Cd} = 0$ V<br>$v_{Coss} = V_i$ | $I_{Lm}$ → constant<br>L → deliver energy<br>$D_1$ and $D_2$ → positive biased | S is turned-on |
| $t_{ON\text{-}T}$<br>$(t_5 – t_6)$ | | S is turned-on | Cd capacitor charged with negative polarity | $i_{Lksec} = i_L$<br>$v_{Cd} = V_{Cd_min}$ |

The voltage and current waveforms and their expressions are shown by component. For the sake of simplicity, the voltage drop through the leakage inductances is neglected. The following nomenclature rules are considered in the expressions shown below:

- The mean and constant values are denoted with capital letters.
- The time-dependent variables are denoted with lower case letters. Also, the time dependence is denoted by "(t)". Different suffixes are used when each expression applies, see Table II.

Table II. Relationship between the expression suffixes and the corresponding time intervals

| Suffix | Switching intervals | Time intervals |
|---|---|---|
| *_ON | $t_{ON}$ | $t \in (t_0, t_1)$ |
| *_OFF-T | $t_{OFF-T}$ | $t \in (t_1, t_2)$ |
| *_OFF | $t_{OFF}$ | $t \in (t_2, t_5)$ |
| *_OFF1 | $t_{OFF1}$ | $t \in (t_2, t_3)$ |
| *_OFF2 | $t_{OFF2}$ | $t \in (t_3, t_4)$ |
| *_OFF3 | $t_{OFF3}$ | $t \in (t_4, t_5)$ |
| *_res | $t_{OFF1}$ and $t_{OFF2}$ | $t \in (t_2, t_4)$ |
| *_ON-T | $t_{ON-T}$ | $t \in (t_5, t_6)$ |

- Output filter inductance (L)

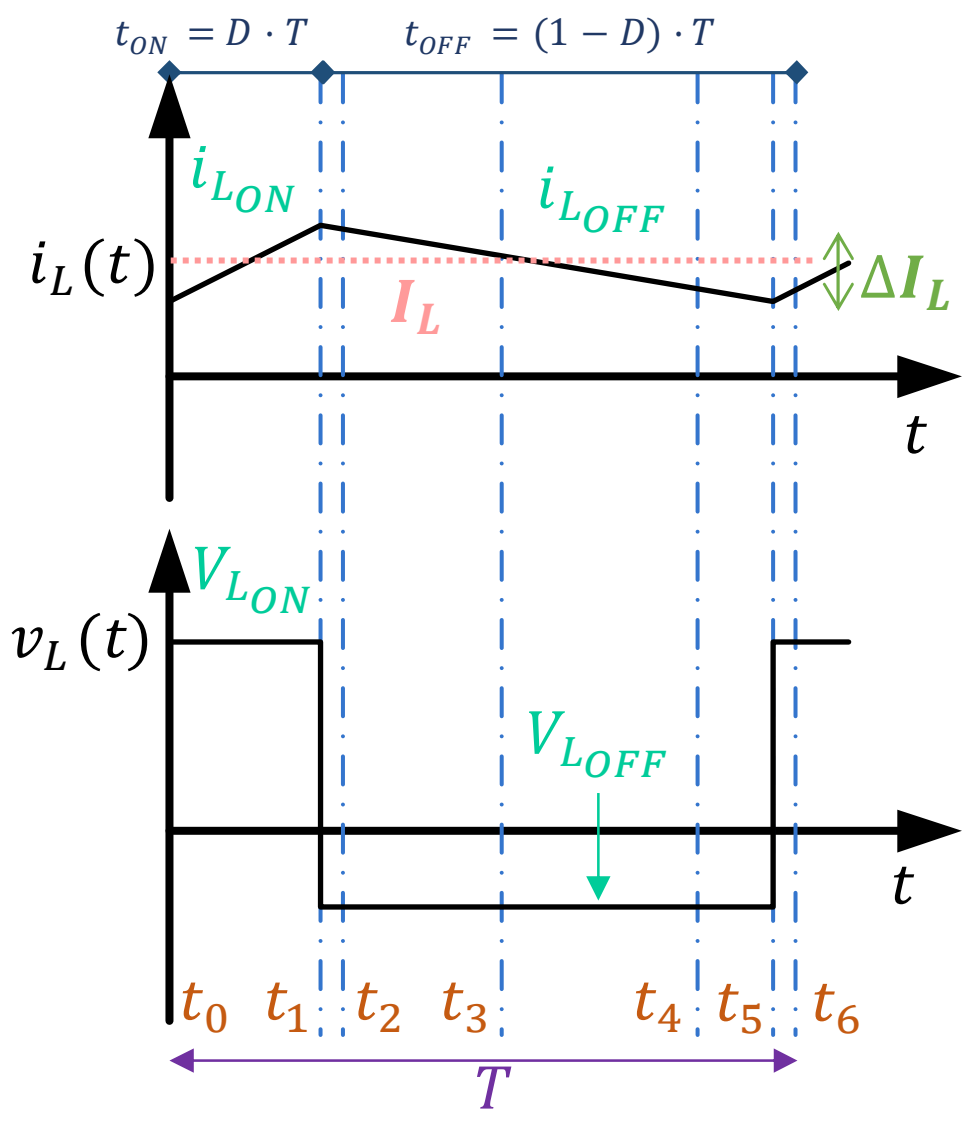

Figure 12. Output inductor current and voltage waveforms, $i_L$ and $v_L$ respectively, during one switching period

$$V_{L_{ON}} = V_i \cdot (1+n) \cdot (1-D) \tag{4}$$

$$V_{L_{OFF}} = -V_i \cdot (1+n) \cdot D \tag{5}$$

$$\Delta I_L = \frac{V_i \cdot (1+n) \cdot (1-D) \cdot D}{L \cdot f_{sw}} \tag{6}$$

$$I_L = I_{string} = \frac{P_o}{V_i \cdot (1+n) \cdot D} \tag{7}$$

$$I_{L_{min}} = I_L - \frac{\Delta I_L}{2} \tag{8}$$

$$i_{L_{ON}}(t) = \frac{V_{L_{ON}}}{L} \cdot t + I_L - \frac{\Delta I_L}{2} \tag{9}$$

$$i_{L_{OFF}}(t) = I_L + \frac{\Delta I_L}{2} - \frac{V_{L_{OFF}}}{L} \cdot (t - t_1) \tag{10}$$

- Autotransformer

Due to the resonant reset, the current and voltage expressions are defined through the following variables:

$$C_{eq} = C_{oss} + C_d \tag{11}$$

$$\omega_0 = 2 \cdot \pi \cdot f_{res} \tag{12}$$

$$\omega_{0_k} = 2 \cdot \pi \cdot f_{res_k} \tag{13}$$

$$L_k = L_{kpri} + \frac{L_{ksec}}{(1+n)^2} \tag{14}$$

Where $f_{res}$ and $f_{resk}$ defined in (2) and (15), respectively, and $L_k$ the leakage inductances referred to the primary side.

$$f_{res_k} \cong \frac{1}{2 \cdot \pi \cdot \sqrt{L_k \cdot (C_d + C_{oss})}} \tag{15}$$

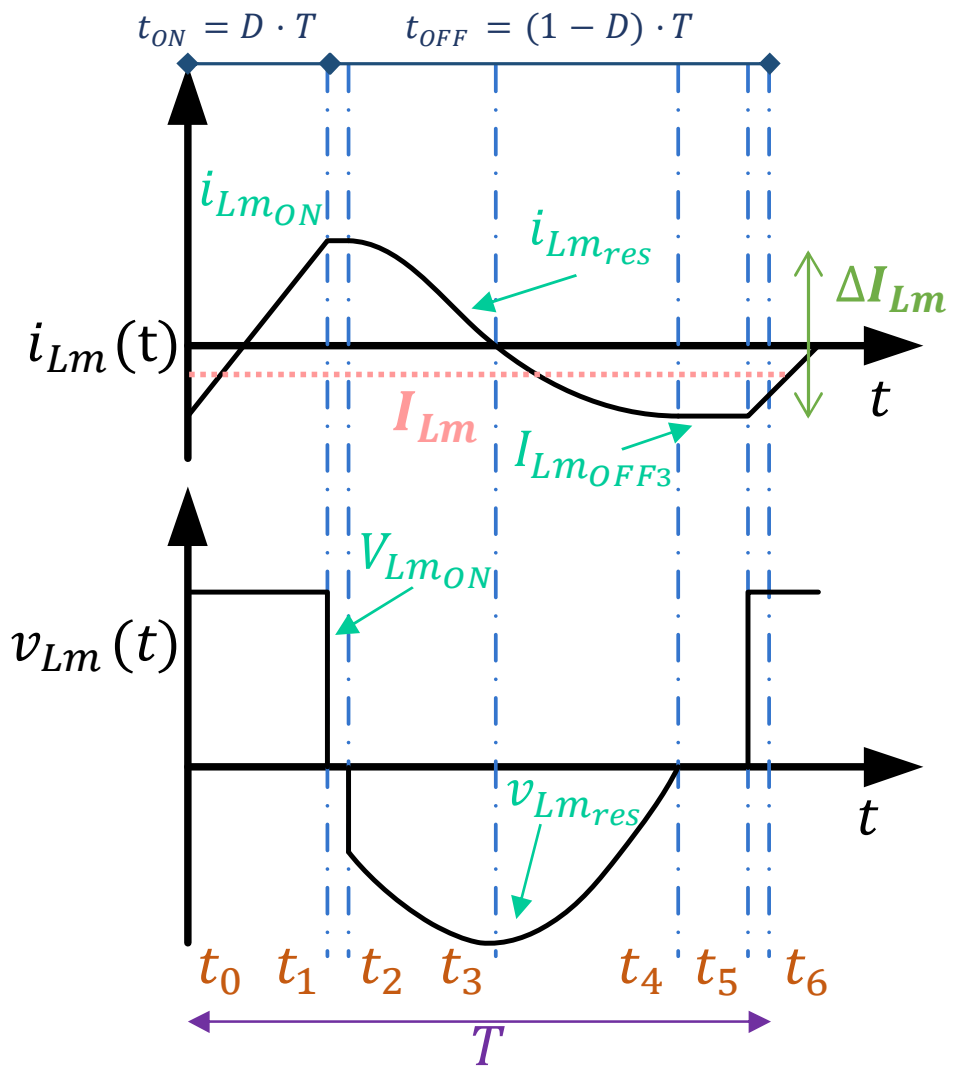

Figure 13. Magnetizing inductor current and voltage waveforms, $i_{Lm}$ and $v_{Lm}$ respectively, during one switching period

$$\Delta I_{Lm} = \frac{V_i \cdot D}{L_m \cdot f_{sw}} \tag{16}$$

$$I_{Lm_{max}} = I_{Lm}(t_1) = I_{Lm_{min}} + \Delta I_{Lm} \tag{17}$$

$$I_{Lm} = I_{Lm_{min}} + \frac{\Delta I_{Lm}}{2} \tag{18}$$

$$V_{Lm_{ON}} = V_i \tag{19}$$

$$v_{Lm}(\mathrm{t}_2) = -v_{Cd}(t_2) \tag{20}$$

$$v_{Lm_{res}}(\mathrm{t}) = v_{Cd_{res}}(t) \tag{21}$$

$$i_{Lm_{ON}}(t) = \frac{V_{Lm_{ON}}}{L_m} \cdot t - \frac{\Delta I_{Lm}}{2} \tag{22}$$

$$i_{Lm_{res}}(t) = I_{Lm_{max}} \cdot \cos(\omega_0 \cdot (t - t_2)) \tag{23}$$

$$I_{Lm_{OFF3}} = I_{Lm_{min}} \tag{24}$$

- Cd capacitor

$$i_{Cd_{res}}(t) = i_{Lm_{res}}(t) \tag{25}$$

$$I_{Cd_{max}} = I_S(\mathrm{t}_1) \tag{26}$$

$$I_{Cd_{min}} = I_{L_{min}} \tag{27}$$

$$v_{Cd_{OFF-T}}(t) \cong V_{Cd_{min}} + I_S(\mathrm{t}_1) \cdot \sqrt{\frac{L_k}{C_d}} \cdot \sin\left(\omega_{0_k} \cdot (t - t_1)\right) \tag{28}$$

$$v_{Cd_{res}}(t) \cong V_{Cd_{OFF-T}}(t_2) + I_{Lm_{max}} \cdot \sqrt{\frac{L_m}{C_d}} \cdot \sin\left(\omega_0 \cdot (t - t_2)\right) \tag{29}$$

$$V_{Cd_{min}} \cong -\left(I_{L_{min}} - \frac{I_{Lm_{min}}}{1+n}\right) \cdot \sqrt{\frac{L_k}{C_d}} \cdot \sin\left(\omega_{0_k} \cdot (t_6 - t_5)\right) \quad (30)$$

$$V_{Cd_{max}} = V_{Cd_{res}}(t_3) \quad (31)$$

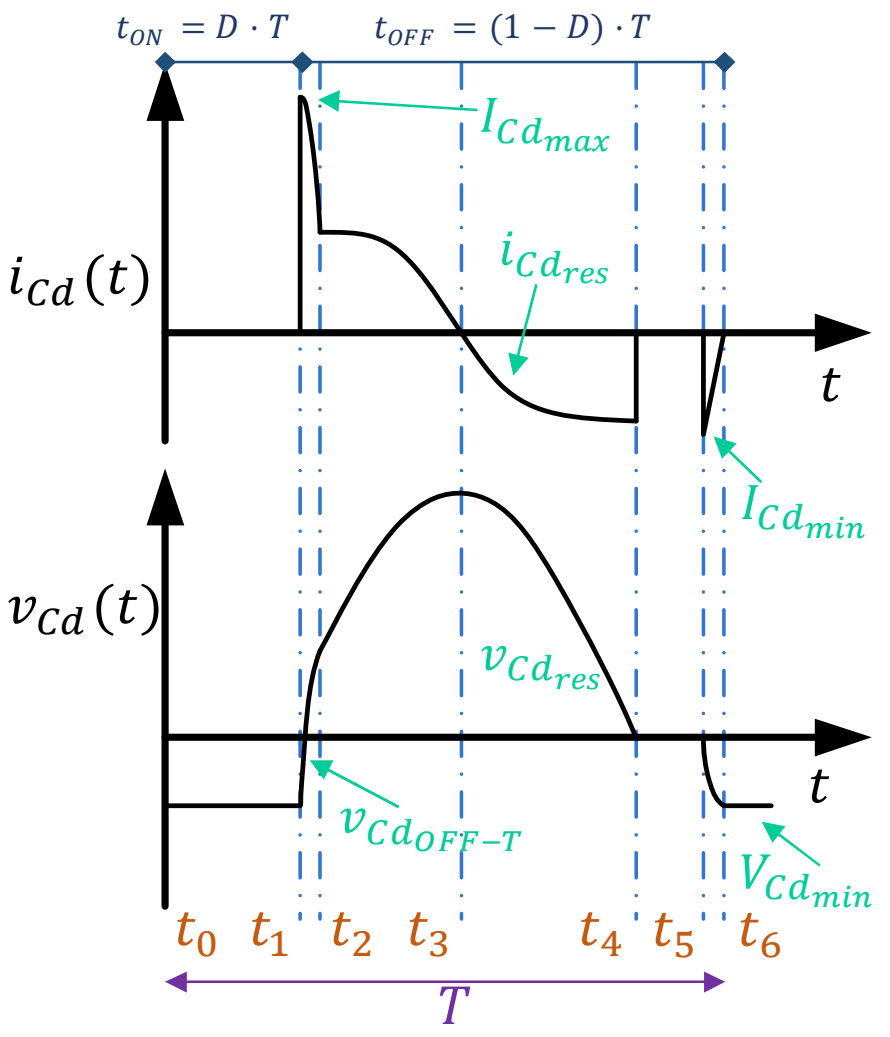

Figure 14. $C_d$ capacitor current and voltage waveforms, $i_{Cd}$ and $v_{Cd}$ respectively, during one switching period

- MOSFET (S)

$$i_{S_{ON}}(t) = i_{L_{ON}}(t) \cdot (1+n) + i_{Lm_{ON}}(t) \quad (32)$$

$$I_S \cong I_L \cdot (1+n) \cdot D + I_{Lm} \quad (33)$$

$$v_{DS_{OFF-T}}(t) = V_i + v_{Cd_{OFF-T}}(t) \quad (34)$$

$$v_{DS_{res}}(t) = V_i + v_{Cd_{res}}(t) \quad (35)$$

$$V_{DS_{OFF3}}(t) = V_{\mathrm{i}} \quad (36)$$

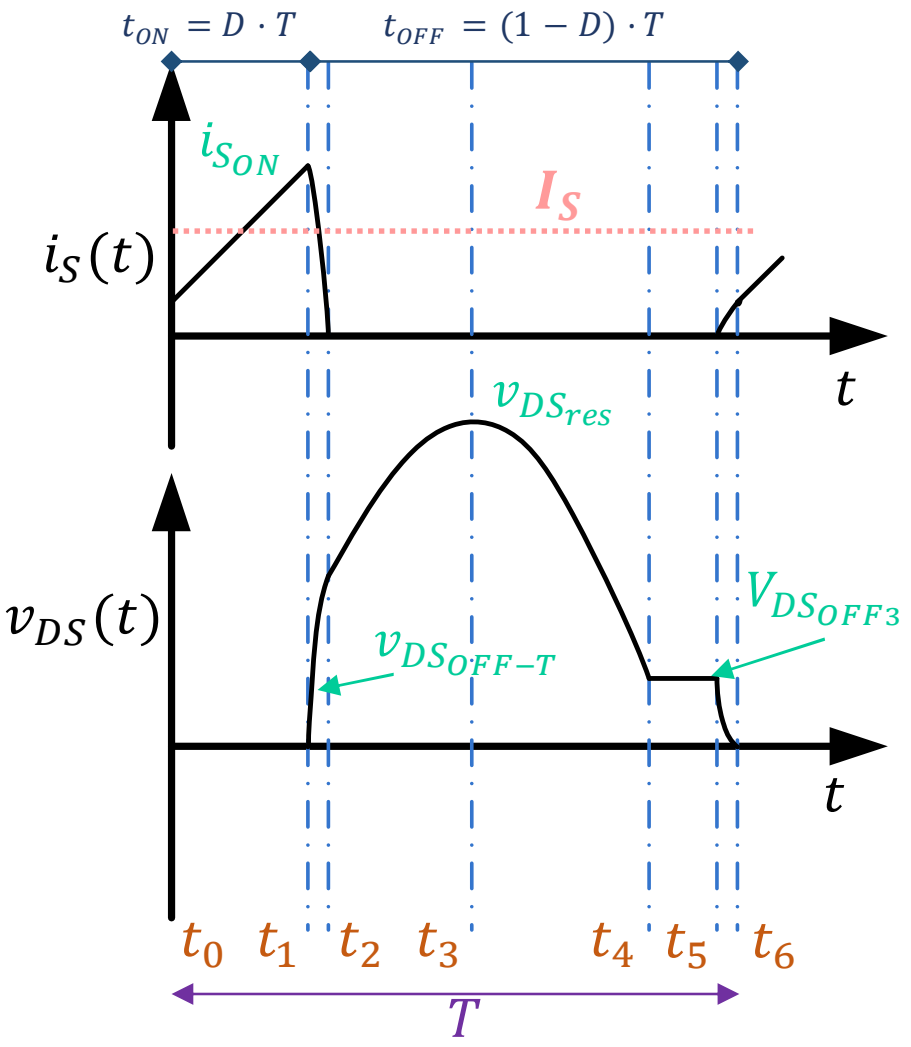

Figure 15. MOSFET current and voltage waveforms, $i_{DS}$ and $v_{DS}$ respectively, during one switching period

- D₁ diode

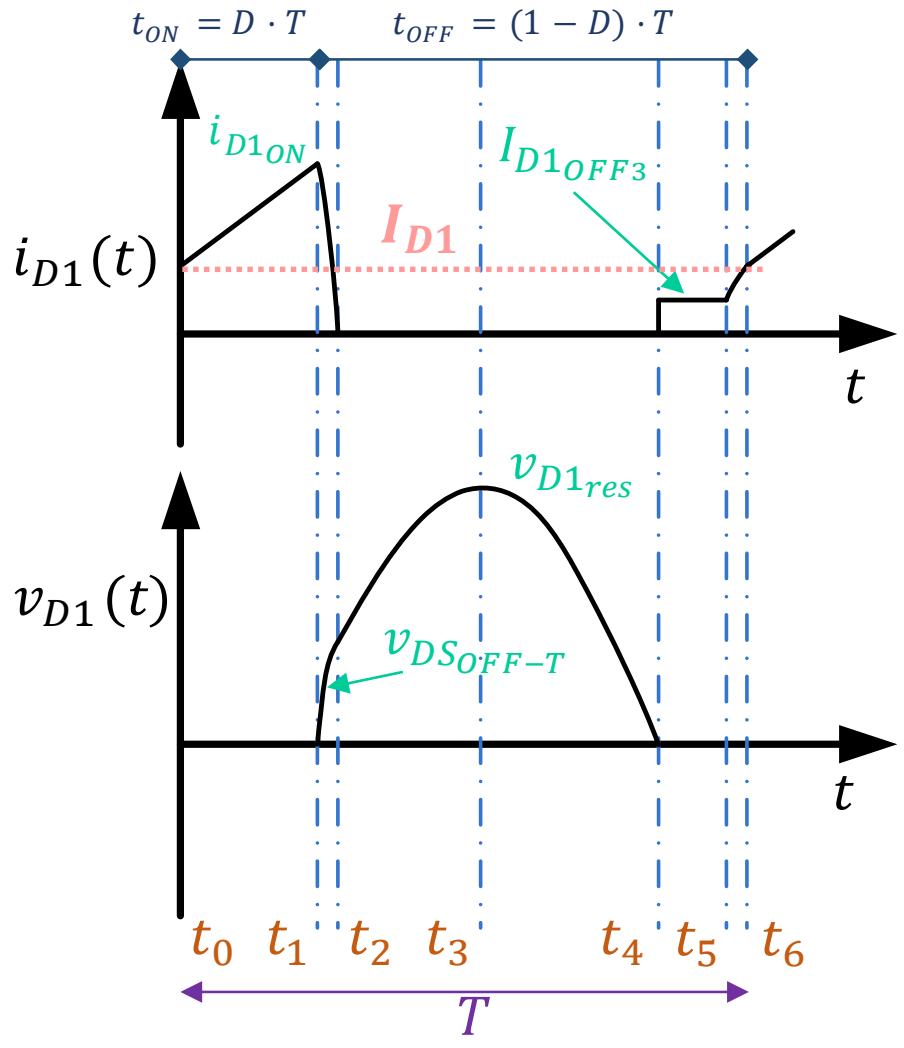

Figure 16. $D_1$ diode current and voltage waveforms, $i_{D1}$ and $v_{D1}$ respectively, during one switching period

$$i_{D1_{ON}}(t) = i_{L_{ON}}(t) \tag{37}$$

$$I_{D1_{OFF3}} = -\frac{I_{Lm_{OFF3}}}{1+n} \tag{38}$$

$$I_{D1} \cong I_L \cdot D + I_{D1_{OFF3}} \cdot \frac{(t_5 - t_4)}{T} \tag{39}$$

$$v_{D1_{OFF-T}}(t) = v_{Cd_{OFF-T}}(t) \cdot (1+n) \tag{40}$$

$$v_{D1_{res}}(t) = v_{Cd_{res}}(t) \cdot (1+n) \tag{41}$$

- D2 diode

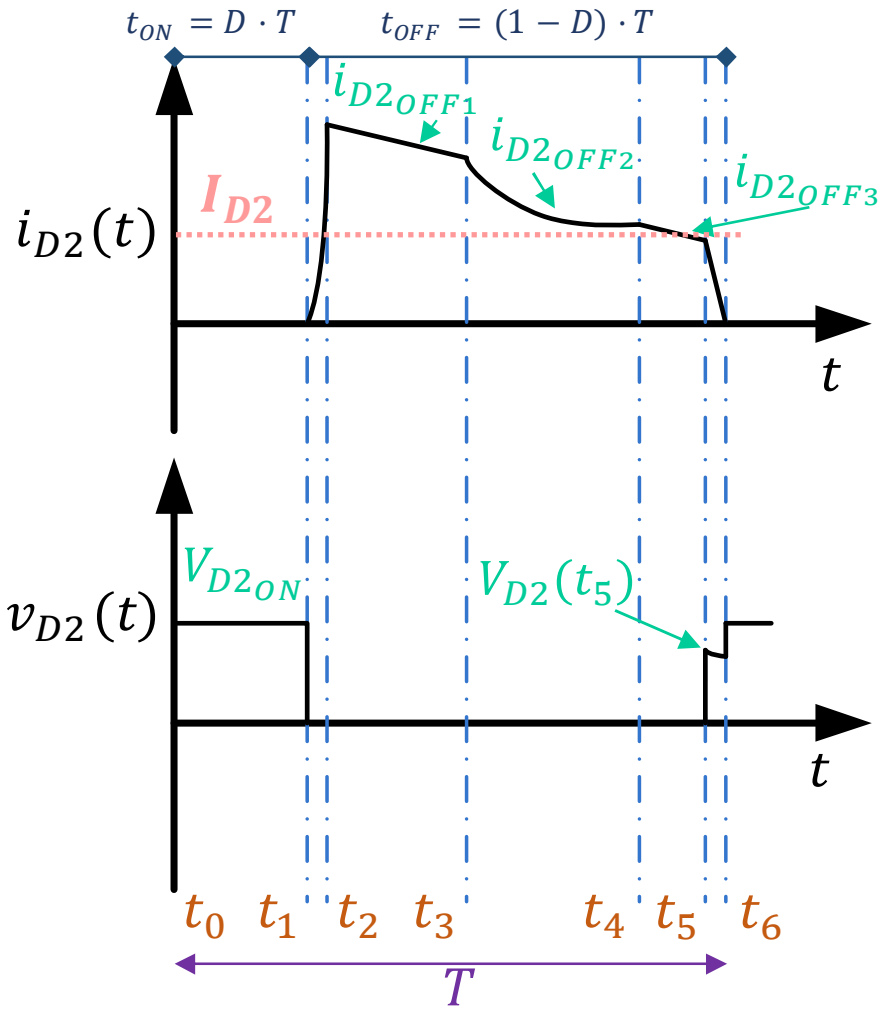

Figure 17. $D_2$ diode current and voltage waveforms, $i_{D2}$ and $v_{D2}$ respectively, during one switching period

$$i_{D2_{OFF1}}(t) = i_{L_{OFF}}(t) \tag{42}$$

$$i_{D2_{OFF2}}(t) = i_{L_{OFF}}(t) - i_{Dd2}(t) \tag{43}$$

$$i_{D2_{OFF3}}(t) = i_{L_{OFF}}(t) - I_{D1_{OFF3}} \tag{44}$$

$$I_{D2} \cong I_L \cdot (1-D) - I_{Dd2} - I_{D1_{\mathrm{OFF3}}} \cdot \frac{(t_5 - t_4)}{T} \cong I_L \cdot (1-D) \quad (45)$$

$$V_{D2_{ON}} = V_i \cdot (1+n) \quad (46)$$

- $D_{d1}$ and $D_{d2}$ diodes

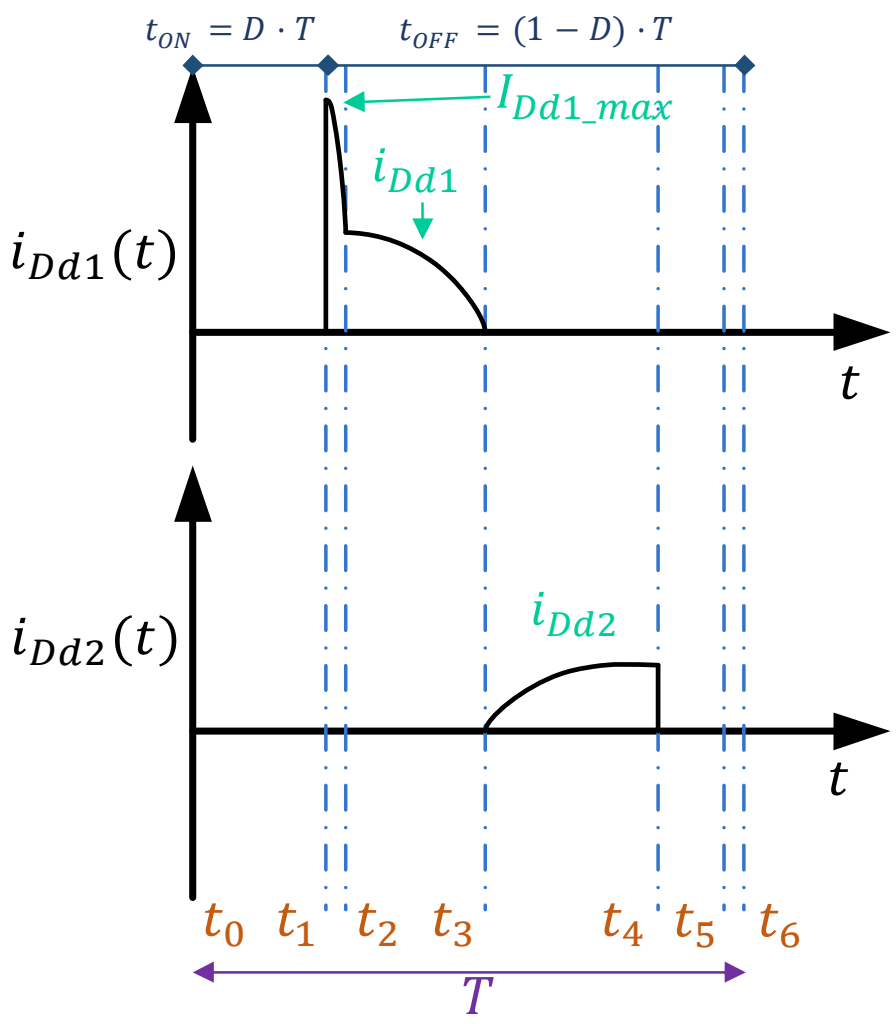

Figure 18. $D_{d1}$ and $D_{d2}$ diodes currents waveforms, $i_{Dd1}$ and $i_{Dd2}$ respectively, during one switching period

$$i_{Dd1}(t) = i_{Cd_{\mathrm{res}}}(t) \quad (47)$$

$$I_{Dd1_{max}} = I_{Cd_{max}} \quad (48)$$

$$i_{Dd2}(t) = -i_{Cd_{\mathrm{res}}}(t) \quad (49)$$

*C. Analysis of the power processing in the autotransformer*

This section includes a detailed analysis of power processing in the AFZ converter autotransformer. The analysis uses the simplified scheme shown in Figure 19. The analysis assumes an ideal efficiency of 100 %, meaning that the input power ($P_i$) is the same as the output power ($P_o$): $P_i = P_o = P_{nom}$

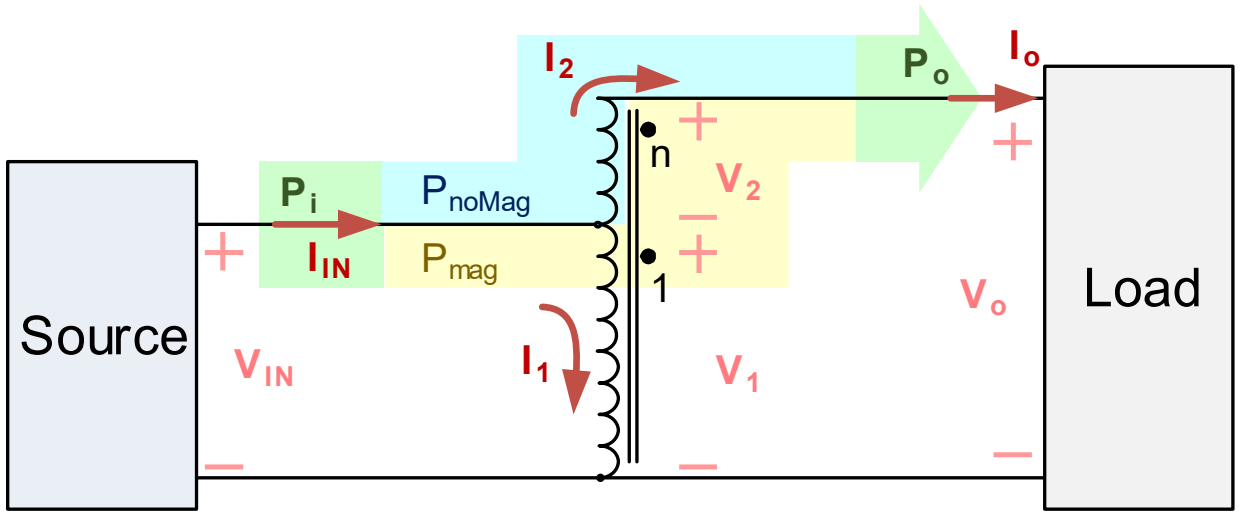

Figure 19. Simplified scheme for the autotransformer power processing analysis in an AFZ converter

The power magnetically processed by the autotransformer is denoted as $P_{mag}$, whereas the not magnetically processed is denoted as $P_{noMag}$. The magnetizing inductance is not included in the autotransformer scheme since it does not influence the energy delivered to the load.

In the AFZ converter, all the energy ($E_{TOT}$) is transferred from the source to the load during the $t_{ON}$ switching interval, see (50).

$$E_{TOT} = V_{IN} \cdot I_{IN} \cdot t_{ON} \quad (50)$$

The input voltage and current can be easily obtained from Figure 19.

$$V_{IN} = V_1 \quad (51)$$

$$I_{IN} = (1+n) \cdot I_2 \quad (52)$$

The energy magnetically processed is defined as:

$$E_{mag} = V_1 \cdot I_1 \cdot t_{ON} = V_2 \cdot I_2 \cdot t_{ON} = \\ = V_{IN} \cdot I_{IN} \cdot t_{ON} \cdot \frac{n}{1+n} = E_{TOT} \cdot \frac{n}{1+n} \quad (53)$$

The magnetically processed power ratio $\left(\frac{P_{mag}}{P_i}\right)$ depends on the autotransformer turns ratio n, see (54).

$$P_{mag} = P_i \cdot \frac{n}{1+n} \quad (54)$$

Therefore, the not magnetically processed power can be defined as:

$$P_{noMag} = P_i \cdot \frac{1}{1+n} \quad (55)$$

Because a part of the power is not processed magnetically compared to an equivalent transformer, the size and power losses of this component can be optimized. Table III shows the normalized magnetically processed and not processed power, for several turns ratio values.

Table III. Normalized $P_{noMag}$ and $P_{mag}$ under several turns ratio values

| **n** | 0.1 | 0.5 | 1 | 1.5 | 2 |
|---|---|---|---|---|---|
| **$P_{noMag}$ (%)** | 90.9 | 66.7 | 50.0 | 40.0 | 33.3 |
| **$P_{mag}$ (%)** | 9.10 | 33.3 | 50.0 | 60.0 | 66.7 |

As can be seen in Table III, lower turn ratio values entail lower magnetically processed power ratios Pmag. Therefore, for the sake of the autotransformer optimization, low n values are desired. However, as it can be seen in the AFZ converter voltage transfer function (3), low turn ratio values limit the voltage step-up. More detail regarding the selection of the optimum turn ratio of the autotransformer is included in Section III.A.2.

### *D. Frequency domain analysis: a small-signal model of the AFZ converter*

#### *1. Theoretical analysis*

Knowing the dynamic performances of the converter is essential for the compensator design, which is implemented in almost every converter. The AFZ converter small-signal model is obtained through the Absorbed and Injected Current Model, described in detail in [38]. With this modeling technique, the dynamic performances of the converter are obtained through their input and output currents, see Figure 20.

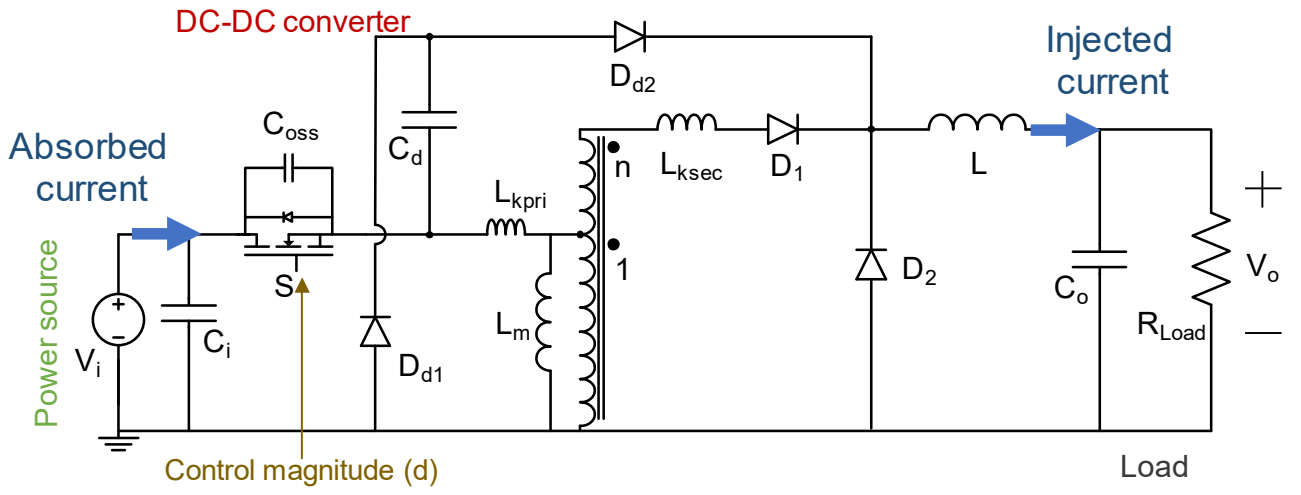

Figure 20. Absorbed and Injected Current Model applied to the AFZ converter

It can be noted that the load includes not only $R_{Load}$ but also the output filter capacitor $C_o$. The parallel association of both $C_o$ and $R_{Load}$ is denoted as $Z_p$ (s). Where $s = j \cdot 2 \cdot \pi \cdot f$.

In this paper, only the output voltage-duty cycle small-signal transfer function, $G_{vd}(s)$, the audio susceptibility, $G_{vv}(s)$, and the output impedance, $Z_o(s)$, are calculated. Because they are all related to the output voltage, they are all calculated from the injected current.

The expression of the output inductor averaged current, for a given frequency, is shown in (56).

$$I_L(s) = \frac{(1+n) \cdot V_i \cdot D - V_o}{Z_L(s)} \quad (56)$$

Where $Z_L(s) = s \cdot L$. After the linearization and perturbation on its operating point, the previous expression leads to:

$$\hat{\imath}_L(s) = \frac{1}{Z_L(s)} \cdot \left[(1+n) \cdot V_i \cdot \hat{d} - \hat{v}_o + (1+n) \cdot D \cdot \hat{v}_i\right] \quad (57)$$

Where the superscript "^" refers to small-signal variables.

From the expressions (57) and (58), the desired small-signal transfer functions can be calculated from the block diagram shown in Figure 21.

$$\hat{\imath}_L(s) = \frac{\hat{v}_o}{Z_p(s)} \quad (58)$$

From the block diagram, the small-signal disturbed output inductance current can be expressed as:

$$\hat{\imath}_L(s) = A(s) \cdot \hat{d} - B(s) \cdot \hat{v}_o + C(s) \cdot \hat{v}_i \quad (59)$$

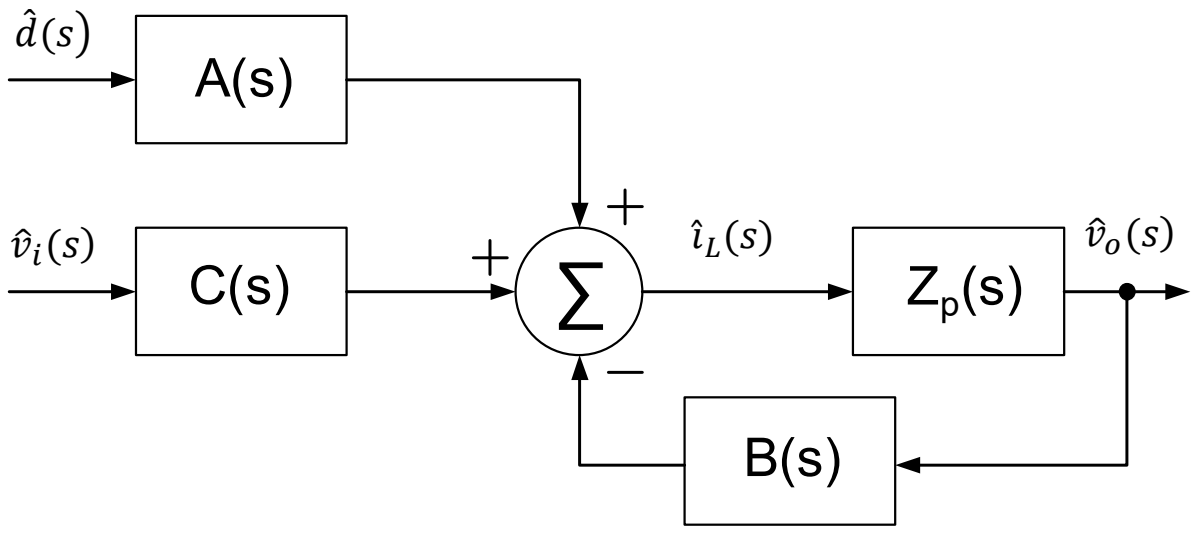

Figure 21. Small-signal block diagram

Comparing the expressions (57) and (58) with the expression (59), the A(s), B(s), and C(s) specific values for the AFZ converter are obtained, see Table IV.

Table IV. Expressions of the AFZ converter small-signal blocks

| Small signal block | Expression |
|---|---|
| $A(s)$ | $\frac{(1+n) \cdot V_i}{Z_L(s)}$ |
| $B(s)$ | $\frac{1}{Z_L(s)}$ |
| $C(s)$ | $\frac{(1+n) \cdot D}{Z_L(s)}$ |
| $Z_L(s)$ | $s \cdot L$ |
| $Z_p(s)$ | $\frac{R_{Load}}{1 + s \cdot C_o \cdot R_{Load}}$ |

The desired small-signal transfer functions can be now defined as:

$$G_{vd}(s) = \frac{\hat{v}_o}{\hat{d}} = \frac{A(s) \cdot Z_p(s)}{1 + B(s) \cdot Z_p(s)} = \\ = (1+n) \cdot V_i \cdot \frac{\omega_o^2}{s^2 + s \cdot \frac{1}{R_L \cdot C_o} + \omega_o^2} \quad (60)$$

$$G_{vv}(s) = \frac{\hat{v}_o}{\hat{v}_i} = \frac{C(s) \cdot Z_p(s)}{1 + B(s) \cdot Z_p(s)} = \\ = (1+n) \cdot D \cdot \frac{\omega_o^2}{s^2 + s \cdot \frac{1}{R_L \cdot C_o} + \omega_o^2} \tag{61}$$

$$Z_o(s) = \frac{\hat{v}_o}{\hat{\imath}_o} = \frac{Z_p(s)}{1 + B(s) \cdot Z_p(s)} = \\ = \frac{1}{C_o} \cdot \frac{s}{s^2 + s \cdot \frac{1}{R_L \cdot C_o} + \omega_o^2} \tag{62}$$

Where A(s), B(s), C(s), and $Z_p(s)$ expressions are detailed in Table IV.

### 2. *Simulation validation*

The theoretical small-signal analysis developed in the previous section is verified comparing the frequency domain representation of the expressions (60) – (62) with simulation results, obtained in PSIM®.

The values employed for this comparison are summarized in Table V.

Table V. Parameters of the AFZ converter for a non-shaded PV panel in Scenario 1

| Parameter | Definition | Value |
|---|---|---|
| $f_{sw}$ | Switching frequency | 50 kHz |
| $V_i$ | Input voltage | 29.3V |
| D | Duty cycle | 0.689 |
| n | Autotransformer turns ratio | 1 |
| L | Output filter inductance | 68 µH |
| $C_o$ | Output filter capacitance | 112 µF |
| $R_L$ | Output load | 7.255Ω |

All the values shown in Table V corresponds to the ones selected in the case of the study section, for a non-shaded PV panel in Scenario 1. The case of study and the converter specifications are defined in Section III.A.

The graphical representation of expressions (60) – (62) is shown in Figure 22 - Figure 24, respectively. In these figures, the theoretical frequency response is depicted with a continuous blue line, whereas the AC simulation results obtained with PSIM are depicted with a dashed orange line.

As can be seen in Figure 22 - Figure 24, the theoretical representation fits the simulation results up to half of the switching frequency, i.e., 25 kHz in this case. It is noteworthy to mention that, for higher frequencies, the simulation results should not be considered.

By analyzing the graphical representation, it can be concluded that the dynamic performances of the AFZ converter are like the Forward converter one. It means that there is no presence of a Right-Half-Plane zero in its small-signal transfer functions. This fact simplifies the compensator design process and allows achieving better dynamic performances, such as faster compensator with higher phase and gain margins, and therefore higher bandwidth in closed loop.

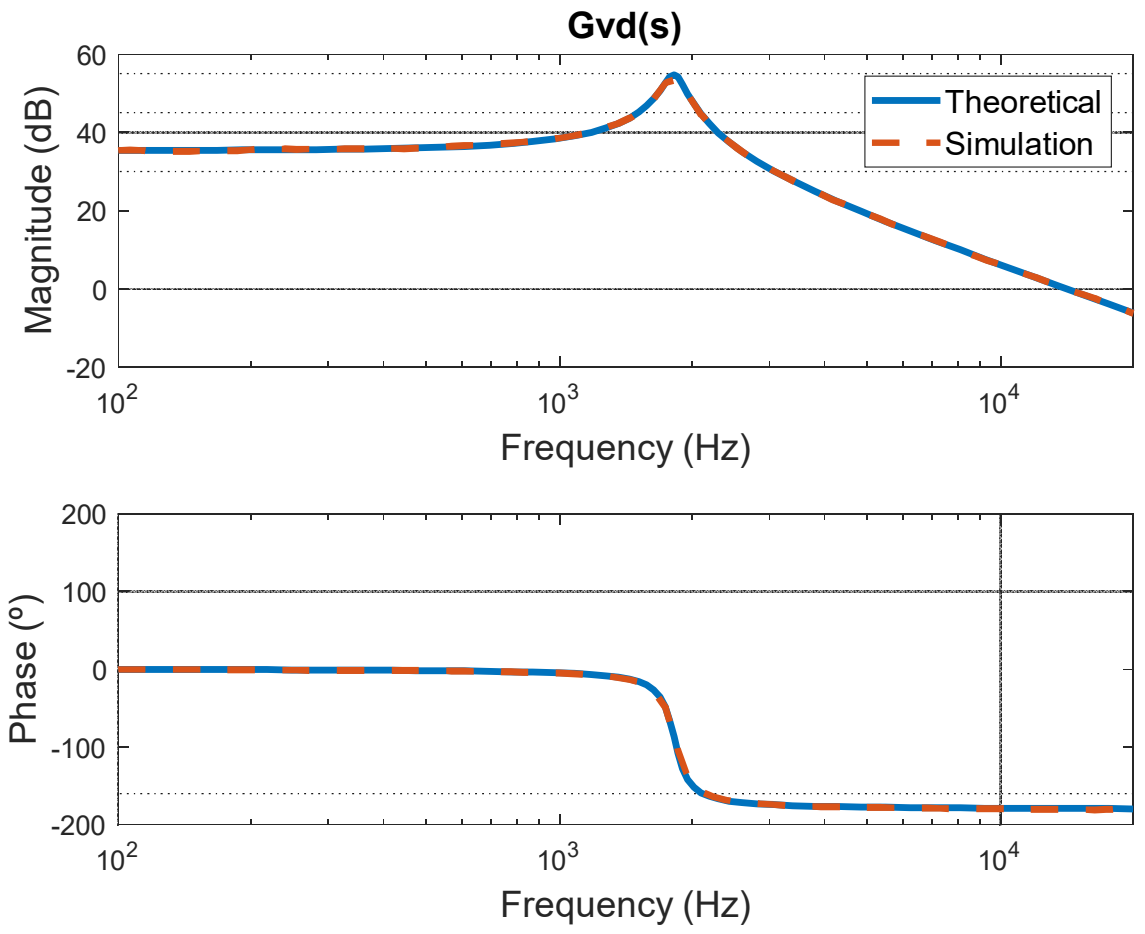

Figure 22. AFZ converter output voltage-duty cycle small-signal transfer function, $G_{vd}(s)$

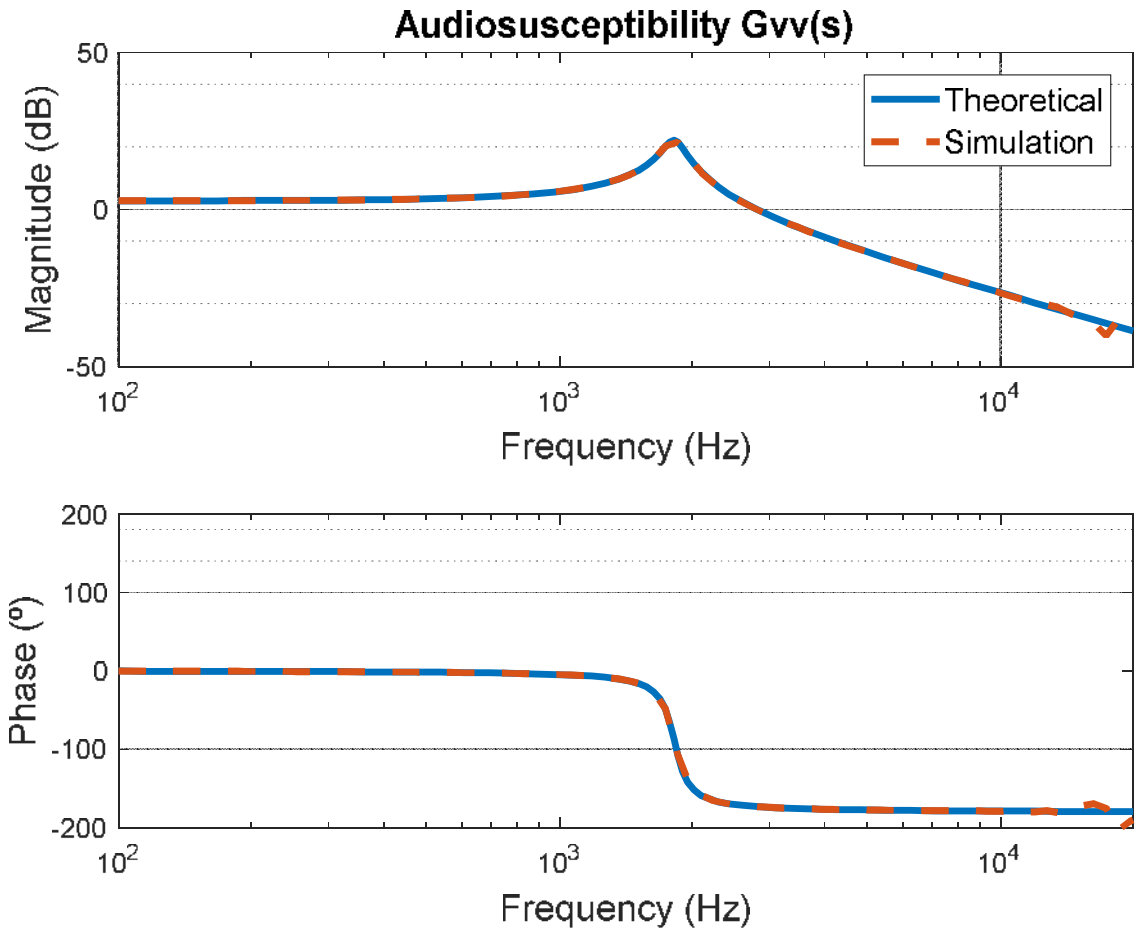

Figure 23. AFZ converter audio susceptibility, $G_{vv}(s)$

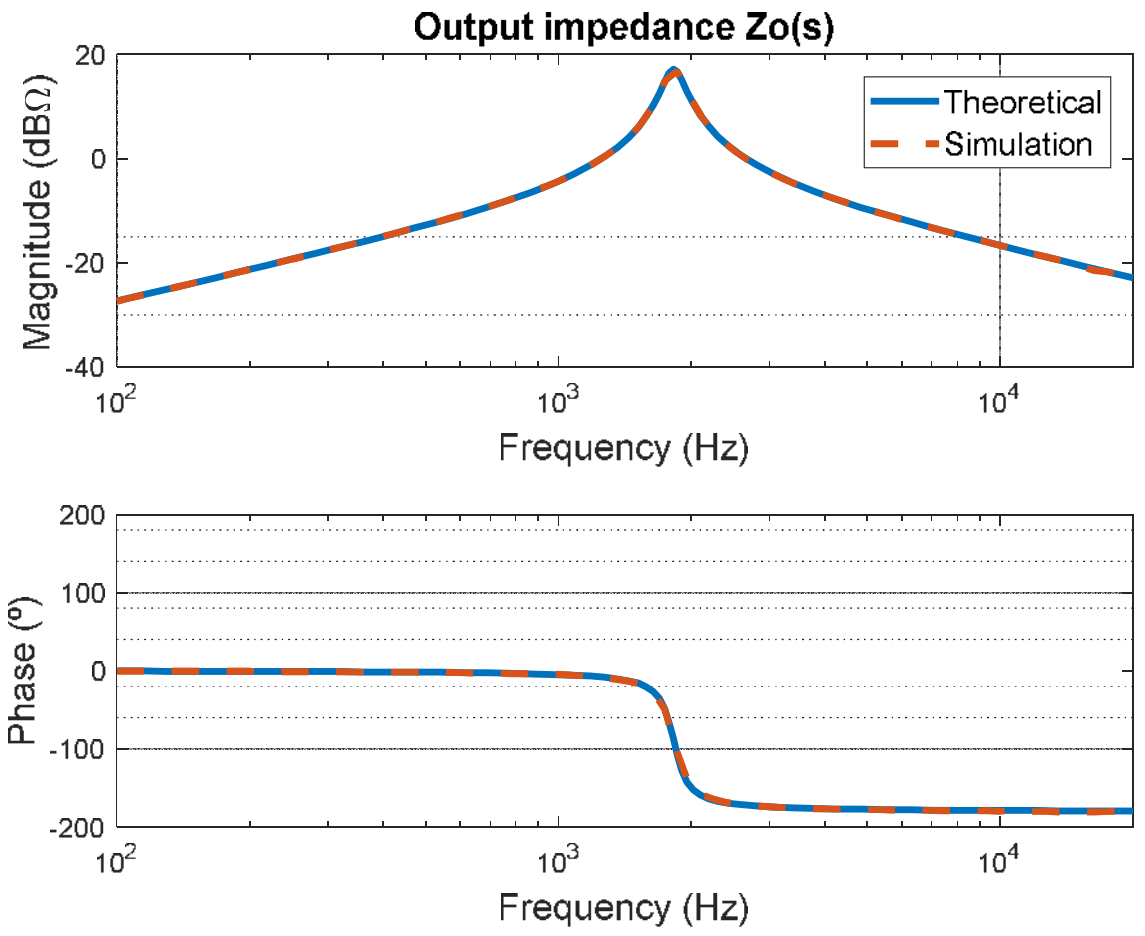

Figure 24. AFZ converter output impedance, $Z_o(s)$

## III. EXPERIMENTAL VERIFICATION

In this section is carried out the design, manufacturing, and implementation of a 225 W AFZ converter prototype. The specific case of study considered in the prototype's design process is detailed in the following section.

### A. Converter design

#### 1. The case of study/application environment

The case of study employed to set the prototype specifications consists of a 100 kW grid-tied PV plant with DMPPT architecture, see Figure 3. As a central inverter, the FREESUN LVT FS0100 inverter is selected [39]. The optimum input voltage of this inverter, which corresponds with the string voltage $V_{string}$, is 600 V. The selected PV panel is the SKJ60P6L, from Siliken [40]. The maximum output power of this PV panel is 225 W.

The installation and the PV panel power are used to determine the number of PV panels needed. In this case, 450 PV panels are required to fulfill the PV installation power specification.

There are several configurations with a different number of PV panels per string and a different number of strings that can be used. The optimum DMPPT architecture considered is one that minimizes the converter input to output voltage step-up while using the minimum number of PV panels that fulfill the installation power generation requirement. This condition is assumed since the efficiency of the converter increases when it is designed for low input to output voltage step-up. The most relevant configurations are listed in Table VI.

As it is highlighted with bold numbers in Table VI, the optimum DMPPT architecture configuration consists of 25 strings with 18 PV panels in series per string, for ideal conditions, and $V_{in}$=29.3 V.

Table VI. Possible PV panel per string and number of strings configurations. The results are sorted from the lowest to the highest total number of PV panels and input to output voltage step-up. $V_{in}$=29.3 V.

| Number of strings | Number of PV panels per string | Total number of PV panels | Converter output voltage (V) | $\lvert\Delta V\rvert$ $(V)$ |
|---|---|---|---|---|
| **25** | **18** | **450** | **33.33** | **4.03** |
| 18 | 25 | 450 | 24.00 | 5.30 |
| 30 | 15 | 450 | 40.00 | 10.70 |
| 45 | 10 | 450 | 60.00 | 30.70 |
| 50 | 9 | 450 | 66.67 | 37.37 |
| 41 | 11 | 451 | 54.55 | 25.25 |
| 35 | 13 | 455 | 46.15 | 16.85 |
| 24 | 19 | 456 | 31.58 | 2.28 |
| 19 | 24 | 456 | 25.00 | 4.30 |
| 38 | 12 | 456 | 50.00 | 20.70 |

Two different scenarios are defined for taking into account the mismatching effect. In one of them, the Scenario 0, no mismatching effect is included, and therefore all the PV panels present the same electrical characteristics. In the other scenario, Scenario 1, 25% of the PV panels are under shadow conditions. It implies that, for Scenario 1, there are two different voltages and power characteristics, one for the non-shaded PV panels and another one for the shaded-PV panels. In reference [3] is explained how a shadow varies the electrical characteristics of a PV panel, depending on how is the shadow level. In this case of study, it is considered a worst-case scenario, in which the shaded PV panels MPP voltage is reduced by a half and the output power by one third, in comparison to the non-shaded PV panels. It should be noted that, for the design of converters, the PV panel output voltage and power are the converter input voltage and power. Two different voltage and power characteristics are therefore defined as inputs for the converter, one for the non-shaded PV panels (29.3V and 225W) and the other one for the shaded ones (15V and 67.5W).

The output voltage range of the converter can be calculated from the string voltage and the number of PV panels per string. It is well known that every PV panel connected to the same string, shares its output current, see Figure 3, therefore

$$I_{string} = \frac{P_{string}}{V_{string}} \tag{63}$$

Where $P_{string}$ and $V_{string}$ the power generated in the string and the voltage of the string, respectively.

In an ideal scenario, where the efficiency of the MIC is 100%, the power delivered by the PV panel is the same as the power delivered to the string by every MIC attached to this PV panel, see Figure 25 and expression (64).

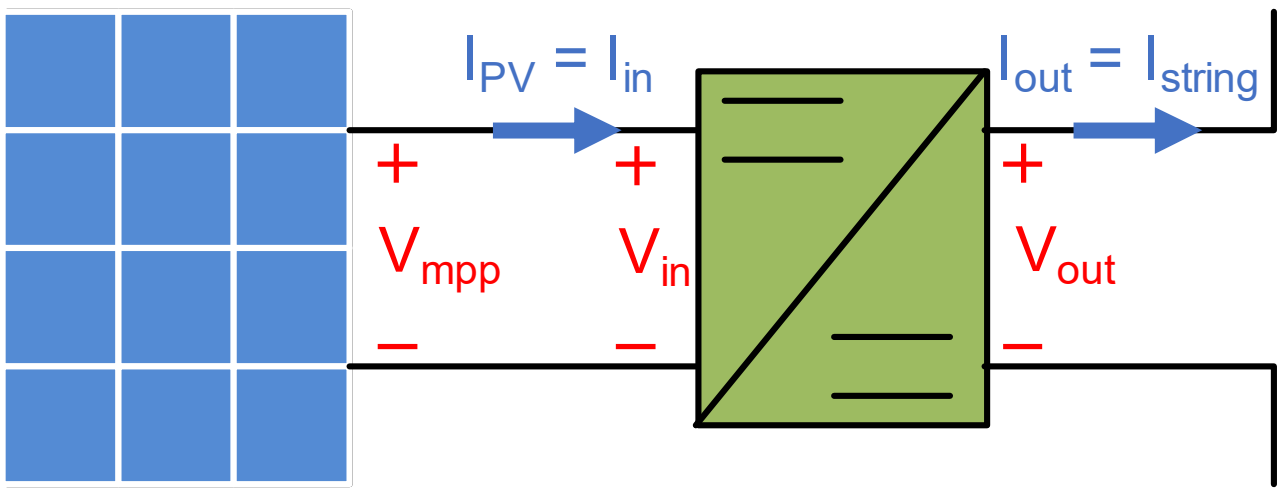

Figure 25. Detail of the connection between the PV panel and the MIC

$$P_{PV} = V_{mpp} \cdot I_{PV} = V_{out} \cdot I_{string} = P_{MIC} \quad (64)$$

Therefore, assuming this statement and taking into account that the inverter controls the string voltage, the output voltage of each converter can be obtained using the full power generated in the string and the power delivered by each PV panel to the string, see (65).

$$V_{out} = \frac{V_{mpp} \cdot I_{PV}}{I_{string}} = \frac{V_{mpp} \cdot I_{PV}}{P_{string}} \cdot V_{string} == \frac{P_{PV}}{P_{string}} \cdot V_{string} \quad (65)$$

Expression (65) can be used to determine the output voltage of the converter as a function of the power delivered to the string.

The parameters obtained from the case of study, for the converters connected to a non-shaded and shaded PV panel, are summarized in Table VII.

Table VII. Case of study parameters summary

| Converter | Parameter | Scenario 0 | Scenario 1 |
|---|---|---|---|
| Non-shaded PV panel | $P_{MIC}$ (W) | 225 | 225 |
| | $V_{in}$ (V) | 29.3 | 29.3 |
| | $V_{out}$ (V) | 33.3 | 40.404 |
| | $I_{string}$ (A) | 6.75 | 5.569 |
| Shaded PV panel | $P_{MIC}$ (W) | N/A | 67.5 |
| | $V_{in}$ (V) | N/A | 15 |
| | $V_{out}$ (V) | N/A | 12.121 |
| | $I_{string}$ (A) | N/A | 5.569 |

It can be noted in Table VII that the converters connected to the non-shaded PV panels have to step-up its output voltage, whereas the ones connected to the shaded PV panels have to step-down its output voltage. This requirement can only be fulfilled by a MIC able to both voltages step-up and step-down.

The specifications for the design of the converter can be obtained, taking into account the parameters shown in Table VII, see Table VIII.

Table VIII. Specifications for the converter design

| Parameter | Specification | Parameter | Specification |
|---|---|---|---|
| $V_i$ *(V)* | [15-29.3] | $D_{max}$ | 0.75 |
| $V_o$ *(V)* | [12-40.4] ±2% | $P_{MICmax}(W)$ | 225 |
| $D_{min}$ | 0 | $P_{MICmin}(W)$ | 60 |

## 2. *Design procedure: the selection of the AFZ converter components*

Once the converter specifications are obtained, see Table VIII, the steps shown below must be followed in the AFZ converter design.

• Selection of a proper turn ratio value. From the autotransformer point of view, the best turn ratio is the minimum that allows a voltage step-up higher enough to reach the maximum output voltage for the maximum duty cycle value, see (3). Fulfilling this thumb rule, the magnetically processed power ratio $\frac{P_{mag}}{P_{out}}$ is minimized, see (54), as well as the autotransformer size.

• Selection of $C_d$. For selecting a proper $C_d$ capacitance, condition (66) must be fulfilled. This condition is obtained from the duty cycle limitation, see (1) and (2).

$$C_d \leq \frac{(1 - D_{max})^2}{(\pi \cdot f_{sw})^2 \cdot L_m} \quad (66)$$

Note that the MOSFET drain-source capacitance $C_{oss}$ is neglected in comparison to $C_d$.

From expression (29) is deduced that low $C_d$ capacitance values imply high voltages. The voltage $V_{Cd}$ is directly related to the voltage stresses withstood by the MOSFET and $D_1$ diode. Therefore, the optimum capacitance value is the maximum that fulfills (66), considering the maximum duty cycle, maximum switching frequency, and maximum magnetizing inductance, in order to reduce the MOSFET and $D_1$ diode stresses.

• Calculation of the most restrictive parameters for the selection of components, considering that the converter is limited to the maximum voltage shown in Table VIII. Equations (4) to (49) are used to determine the electrical stresses withstood by the components. Especial care should be taken with the S MOSFET and $D_1$ diode maximum voltages:

$$V_{DS_max} = V_{DS_{res}}(t_3) \tag{67}$$

$$V_{D1_max} = V_{D1_{res}}(t_3) \tag{68}$$

The current and voltage expressions of the AFZ converter components depend on the following variables: $I_{Lm_{\min}}$, $V_{Cd_{OFF-T}}(t_2)$ and $V_{Cd_{res}}(t_3)$. These variables can be obtained analyzing the energy transfer during the $t_{OFF\text{-}T}$ transition, as well as during the $t_{OFF1}$ and $t_{OFF2}$ subintervals. From the energy analysis, a three equations system is obtained, see expressions (69) to (71). The solutions of this equation system are the aforementioned variables $I_{Lm_{\min}}$, $V_{Cd_{OFF-T}}(t_2)$ and $V_{Cd_{res}}(t_3)$, where $t_2$ and $t_3$ times are defined in Table II.

Note that $I_{Lm_{max}}$ can be directly obtained as a function of $I_{Lm_{\min}}$, as it is described in (17).

• Application of the case of study parameters, detailed in Table VII, to the components current and voltage expressions shown in Section II.B. The solution of the equations system, under Scenario 1 inputs of the case of study, plus the minimum voltage value of the resonant capacitor $V_{Cd_{min}}$ from the expression (30), are summarized in

Table IX. The Scenario 1 inputs are D=0.689, $V_i$=29.3 V, and $I_{string}$=5.569 A.

$$L_{kpri} \cdot \left\{\left[I_{L_{max}} \cdot (1+n) + I_{Lm_{max}}\right]^2 - I_{Lm_{max}}^2\right\} + L_{ksec} \cdot I_{Lm_{max}}^2 = C_{oss} \cdot \left[\left(V_{Cd_{OFF-T}}(t_2) + V_i\right)^2 - V_i^2\right] + C_d \cdot \left(V_{Cd_{OFF-T}}^2(t_2) - V_{Cd_{min}}^2\right) \tag{69}$$

$$L_m \cdot I_{Lm_{max}}^2 = C_{oss} \cdot \left[\left(V_{Cd_{res}}(t_3) + V_i\right)^2 - \left(V_{Cd_{OFF-T}}(t_2) + V_i\right)^2\right] + C_d \cdot \left(V_{Cd_{res}}^2(t_3) - V_{Cd_{OFF-T}}^2(t_2)\right) \tag{70}$$

$$L_m \cdot I_{Lm_{min}}^2 = C_{oss} \cdot \left[\left(V_{Cd_{res}}(t_3) + V_i\right)^2 - V_i^2\right] + C_d \cdot V_{Cd_{res}}^2(t_3) \tag{71}$$

Table IX. Solutions of the equations system under Scenario 1 inputs

| Variable | Value |
|---|---|
| $I_{Lm_{min}}$ | $-0.44\,A$ |
| $V_{Cd_{min}}$ | $-27.27\,V$ |
| $V_{Cd_{OFF-T}}(t_2)$ | $66.18\,V$ |
| $V_{Cd_{res}}(t_3)$ | $156.33\,V$ |

• Other design criteria should be considered, as the output voltage maximum ripple or the limit for ensuring the operation in continuous conduction mode (CCM). For the AFZ converter prototype design, it is considered a maximum output voltage ripple of 2 %. Besides, the output filter inductance has been selected for ensuring the operation of the converter in MCC, whereas the output power is higher than a third of the nominal power.

The main characteristics of the main components selected for the AFZ converter prototype are summarized in Table X.

Table X. Main components of the AFZ converter prototype

| Component | Reference Designator | Main characteristics |
|---|---|---|
| Autotransformer | Self-design (*) | $n = 1$; $L_m = 485\ \mu H$; $L_{Lk} = 820\ nH$; $R_{pri_DC} = 15.5\ m\Omega$; $R_{sec_DC} = 18.3\ m\Omega$ |
| L | 74437529203151 | $L = 150\ \mu H$; $R_{DC} = 30.6\ m\Omega$ |
| $C_i$ | 50SVPF68M | $C = 68\ \mu F$ (x4) |
| $C_o$ | EEHZA1J560P | $C = 56\ \mu F$ (x2) |
| $C_d$ | Ceramic | $C = 11\ nF$ |
| $D_1$ | SCS215AJHR | $V_{max} = 650\ V$; $Q_c = 23\ nC$; $V_f \cong 1.1\ V$ |
| $D_2$ | V40D100C-M3/I | $V_{max} = 100\ V$; $Q_c = 20\ nC$; $V_f \cong 0.3\ V$ |
| $D_{d1}$ and $D_{d2}$ | C3D1P7060Q | $V_{max} = 600\ V$; $Q_c = 4.4\ nC$; $V_f \cong 1\ V$ |
| S | IPB107N20N3 G | $R_{ds_on} = 9.6\ m\Omega$; $Q_g = 65\ nC$ |
| Controller | PIC12F615-I/P | Simple and low cost. |

(*) Manufactured and distributed by RENCO ELECTRONICS, INC., by following our specifications.

The resulting 225 W AFZ converter prototype is shown in Figure 26.

## B. *Time domain measurements*

### 1. *Waveforms measurements*

This section shows the voltage and current waveforms for the most representative components of the AFZ converter, i.e., the S MOSFET and the $D_1$ and $D_2$ diodes. It is important to note that the current is measured using a current-transformer-based sensor. The relationship between the sensed voltage, $v_{sens}$, and the corresponding current is $i = 2 \cdot v_{sens}$. A more detailed description is included in reference [33].

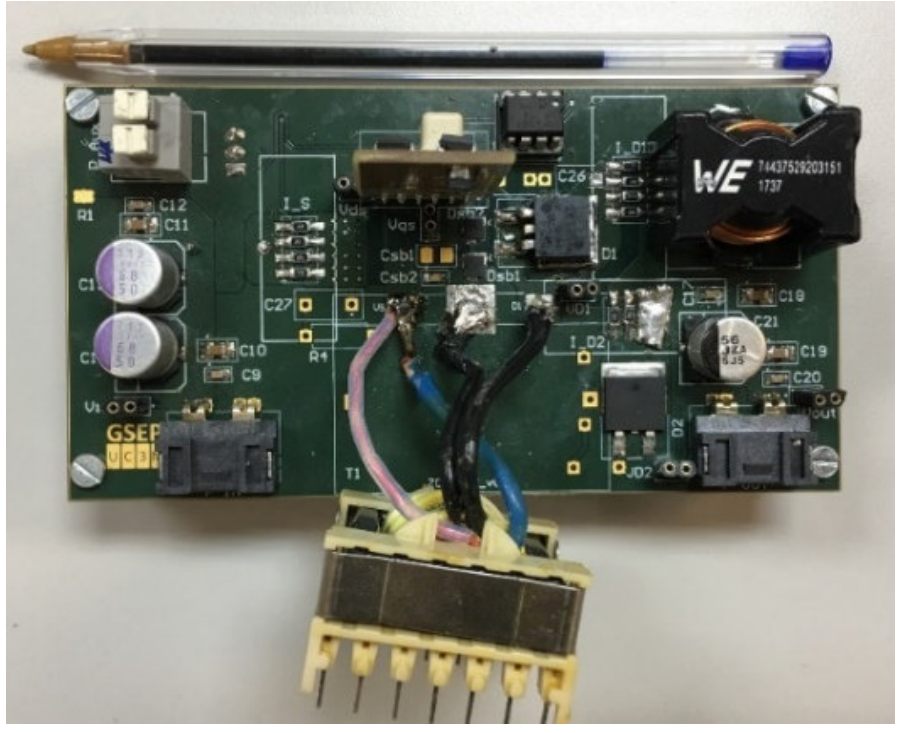

Figure 26. AFZ converter prototype

The test conditions are the same as for the scenarios described in Table VII. Therefore, for each component, three figures are measured, varying the input voltage, the output voltage, and the string current, according to the values defined in Table VII.

Scenario 0 is denoted as S_0, Scenario 1 for PV panels without shadow as S_1, and Scenario 1 for PV panels with shadow as sS_1. The current waveforms are depicted in orange color, whereas the voltage ones are depicted in blue color. All figures include the $t_{ON\text{-}T}$ and $t_{OFF\text{-}T}$ details.

In Figure 27, Figure 28 and Figure 29 are depicted the MOSFET voltage and current waveforms. As can be seen, the waveforms correspond to the theoretical ones, see Figure 15. Regarding the switching transitions, a soft-switching transition quasi-Zero Voltage Switching (ZVS) can be appreciated during the $t_{ON\text{-}T}$ transition, whereas during the $t_{OFF\text{-}T}$ transition, the MOSFET has hard switching, regardless of the measured conditions. The Zeta-type resonant reset subintervals can be appreciated looking at the voltage waveform during the $t_{OFF1}$ and $t_{OFF2}$ subintervals.

In Figure 30 to Figure 32 are depicted the $D_1$ diode waveforms. As can be seen, they match the expected waveforms shown in Figure 16. This diode shows soft-switching characteristics in both transitions, regardless of the operating conditions analyzed, reaching ZVS during the $t_{ON\text{-}T}$ and quasi-Zero Current Switching (ZCS) during the $t_{OFF\text{-}T}$. This kind of switching characteristic reduces the switching losses, increasing the converter efficiency.

The measurements corresponding to the $D_2$ diode are depicted in Figure 33 to Figure 35. Current and voltage waveforms match the ones defined in the theoretical analysis, see Figure 17. The effect of the resonant reset can be appreciated by looking at the current waveform during the $t_{OFF2}$ subinterval. It can be noted the slight variation in the current waveform evolution. It is noteworthy that, waveforms shown in the theoretical analysis section are not at scale. In this case, the $D_2$ diode just shown soft-switching characteristics during the $t_{OFF-T}$ interval, when ZCS is obtained.

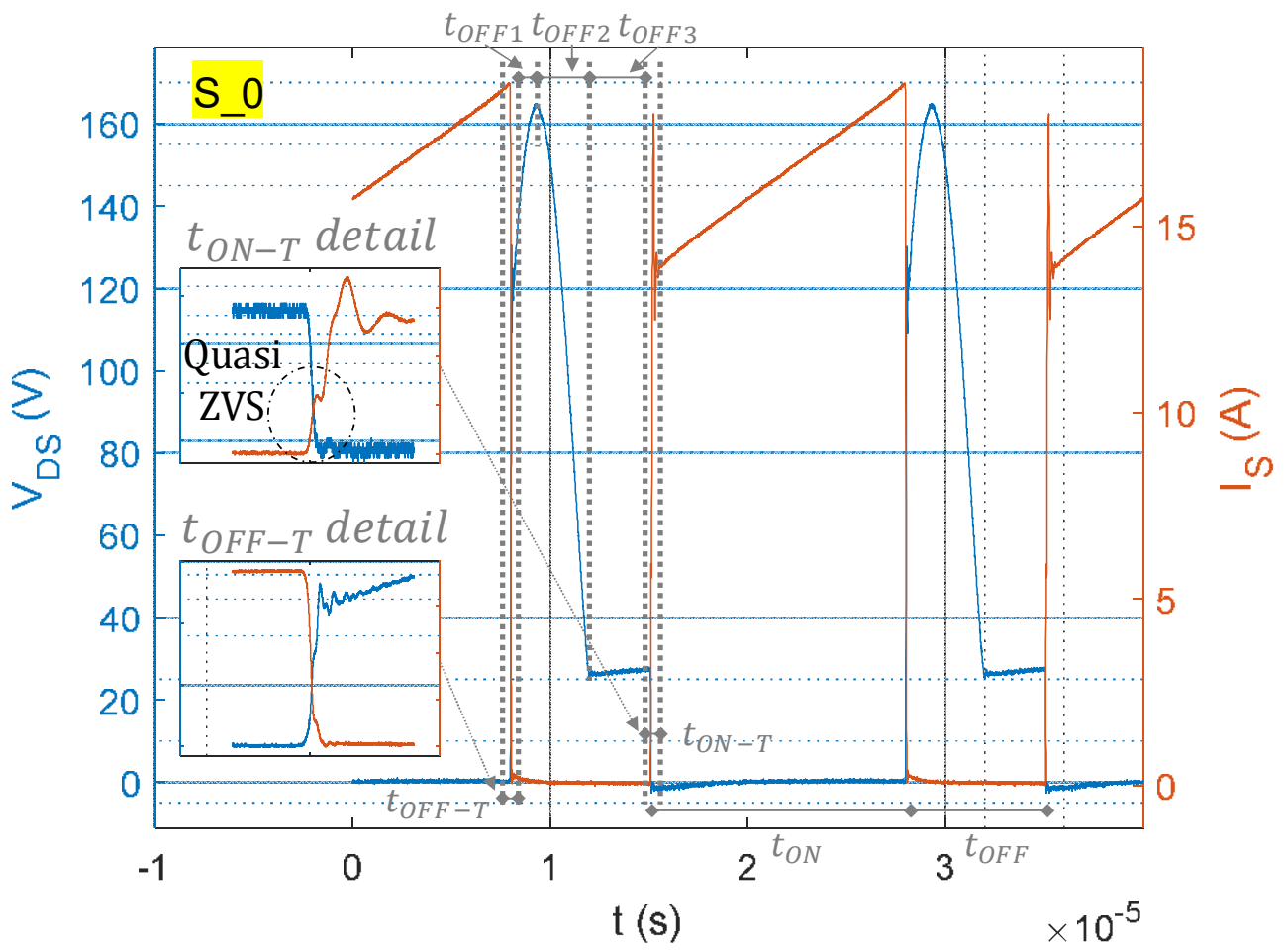

Figure 27. MOSFET waveforms being $V_i = 29.3\ V, V_o = 33.3\ V$ and $I_{string} = 6.75\ A$. $V_{DS}$ in blue color and $I_S$ in orange color

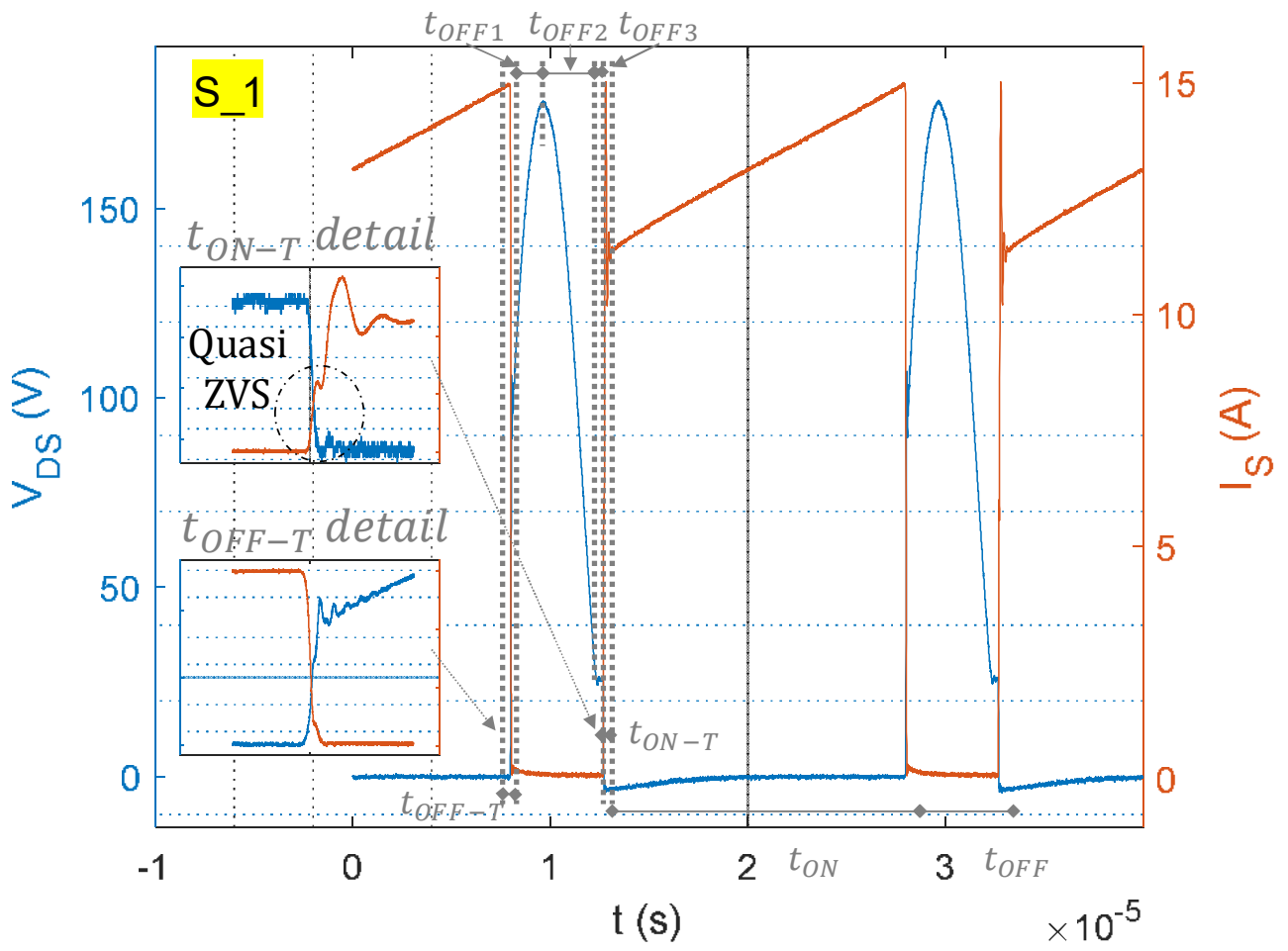

Figure 28. MOSFET waveforms being $V_i = 29.3\ V, V_o = 40.4\ V$ and $I_{string} = 5.569\ A$. $V_{DS}$ in blue color and $I_S$ in orange color

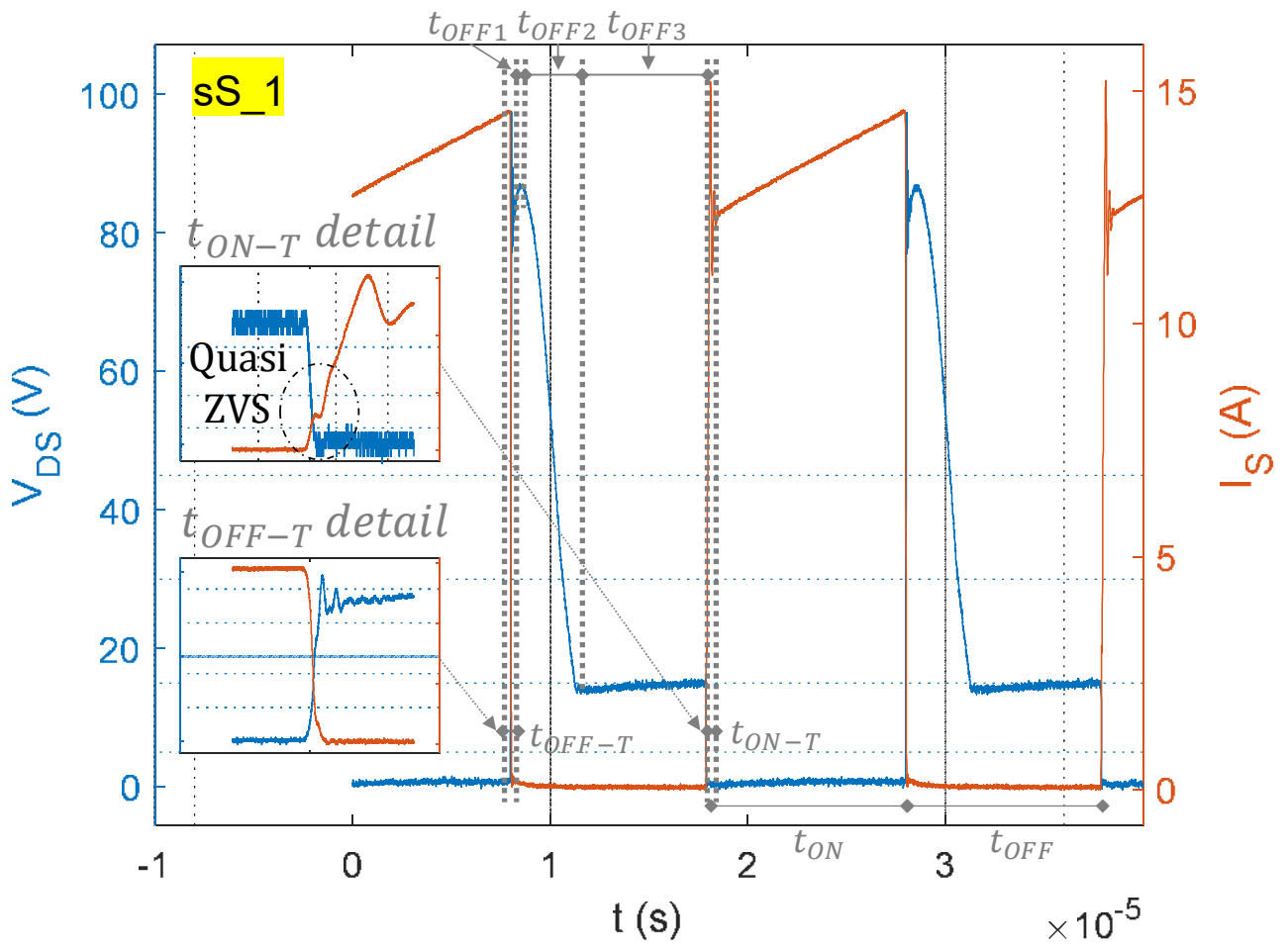

Figure 29. MOSFET waveforms being $V_i = 15\ V, V_o = 12.12\ V$ and $I_{string} = 5.569\ A$. $V_{DS}$ in blue color and $I_S$ in orange color

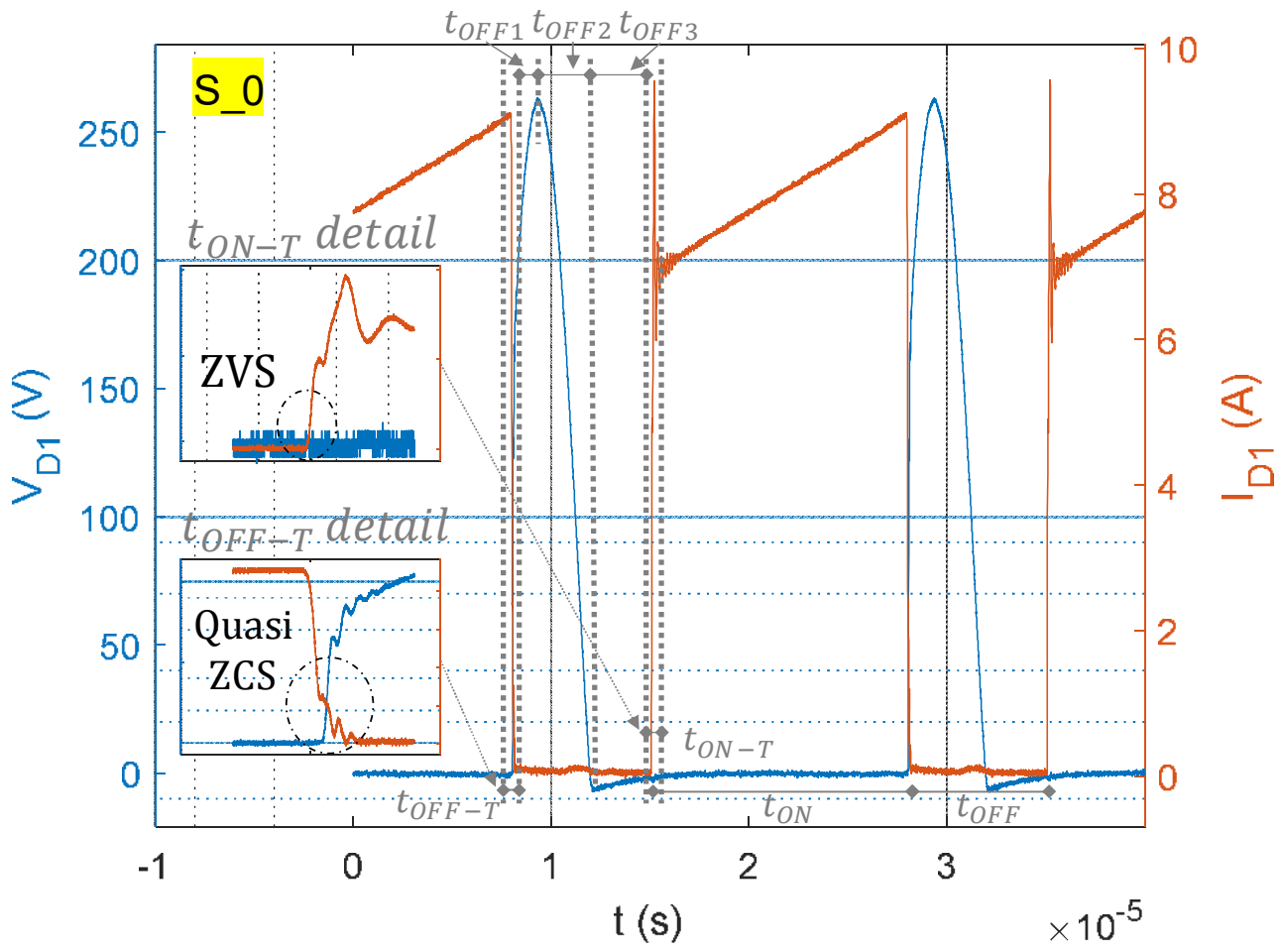

Figure 30. D1 waveforms being $V_i = 29.3\ V, V_o = 33.3\ V$ and $I_{string} = 6.75\ A$. $V_{DS}$ in blue color and $I_S$ in orange color

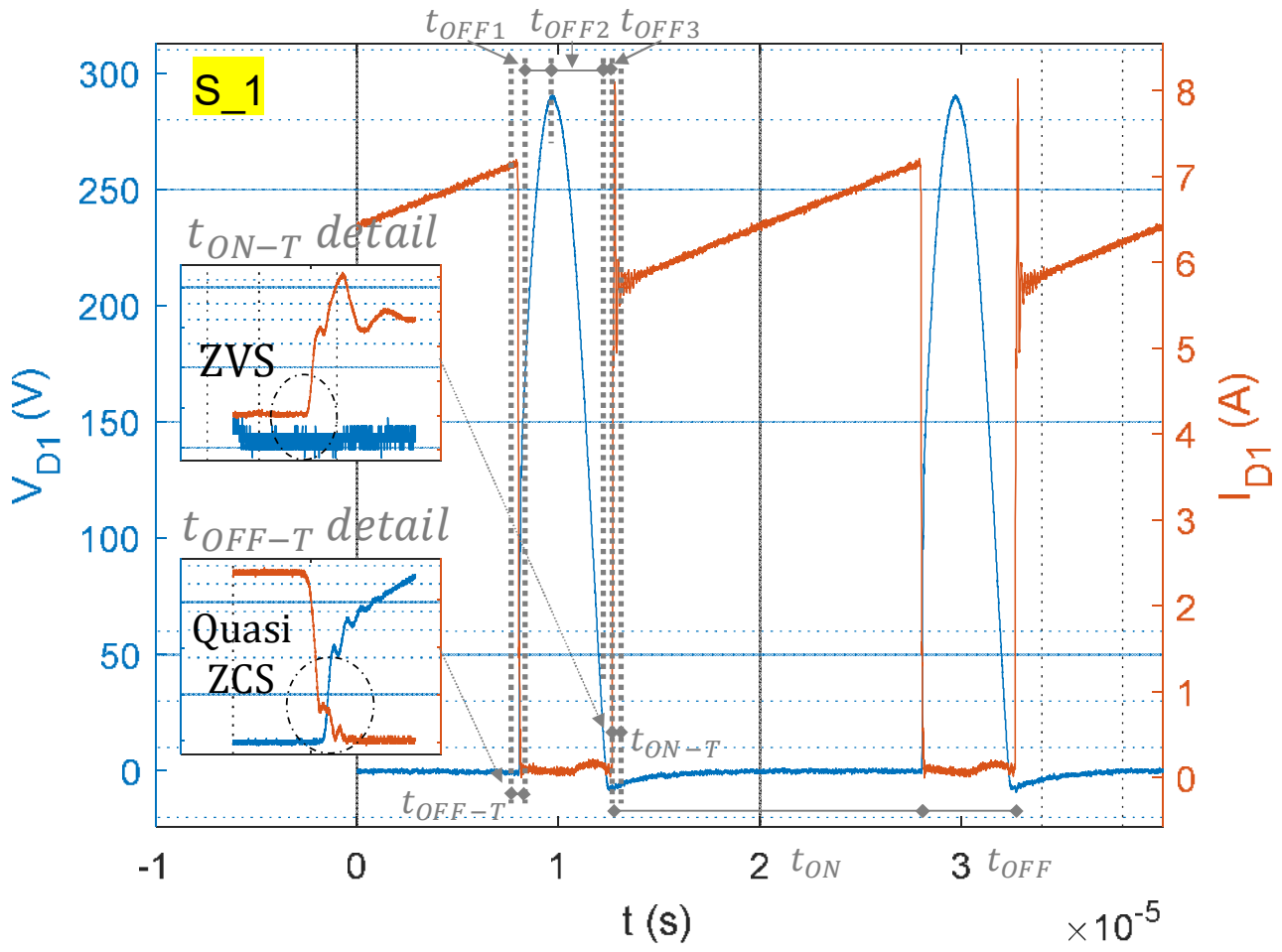

Figure 31. D1 waveforms being $V_i = 29.3\ V, V_o = 40.4\ V$ and $I_{string} = 5.569\ A$. $V_{DS}$ in blue color and $I_S$ in orange color

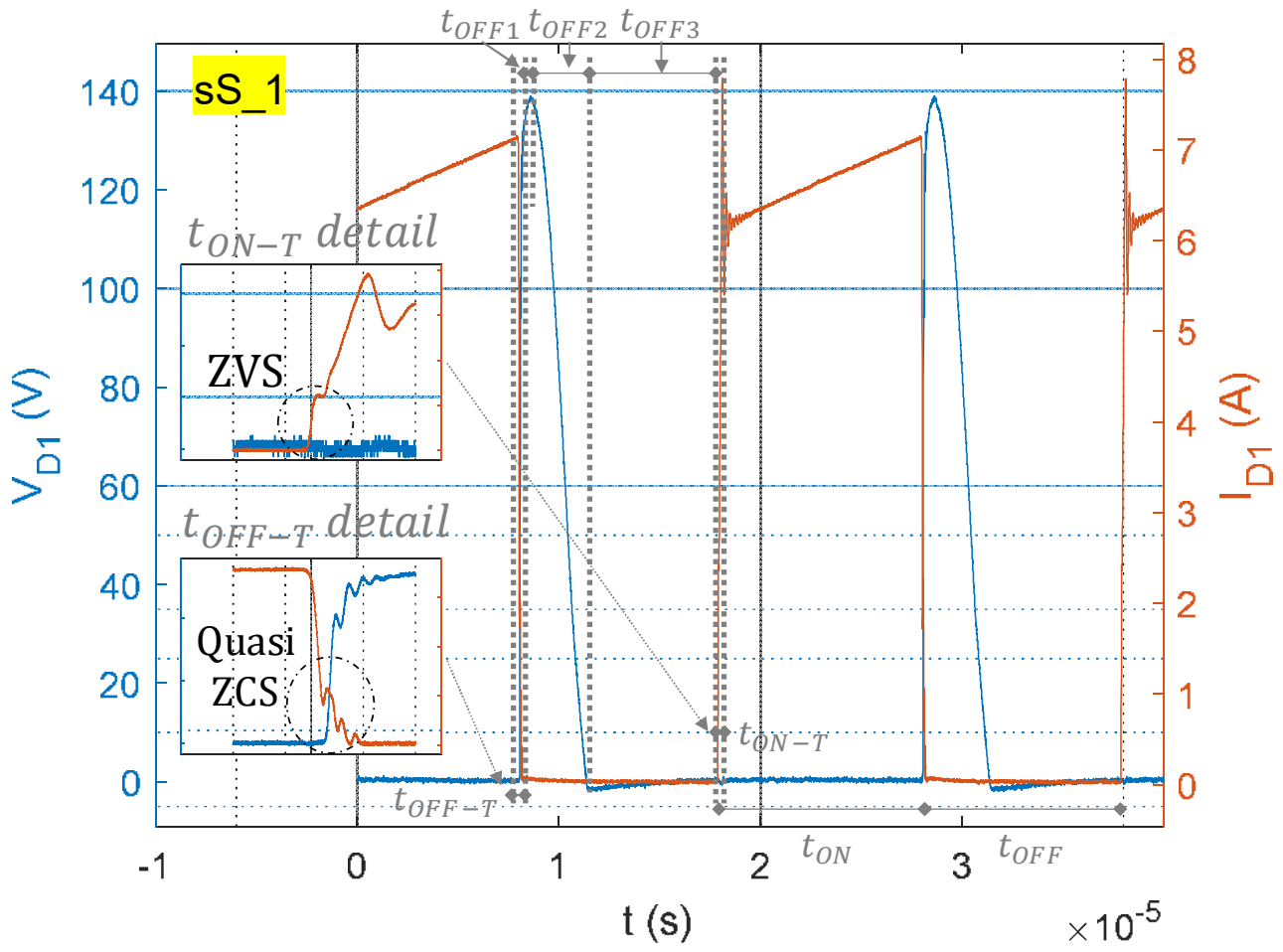

Figure 32. D1 waveforms being $V_i = 15\ V, V_o = 12.12\ V$ and $I_{string} = 5.569\ A$. $V_{DS}$ in blue color and $I_S$ in orange color

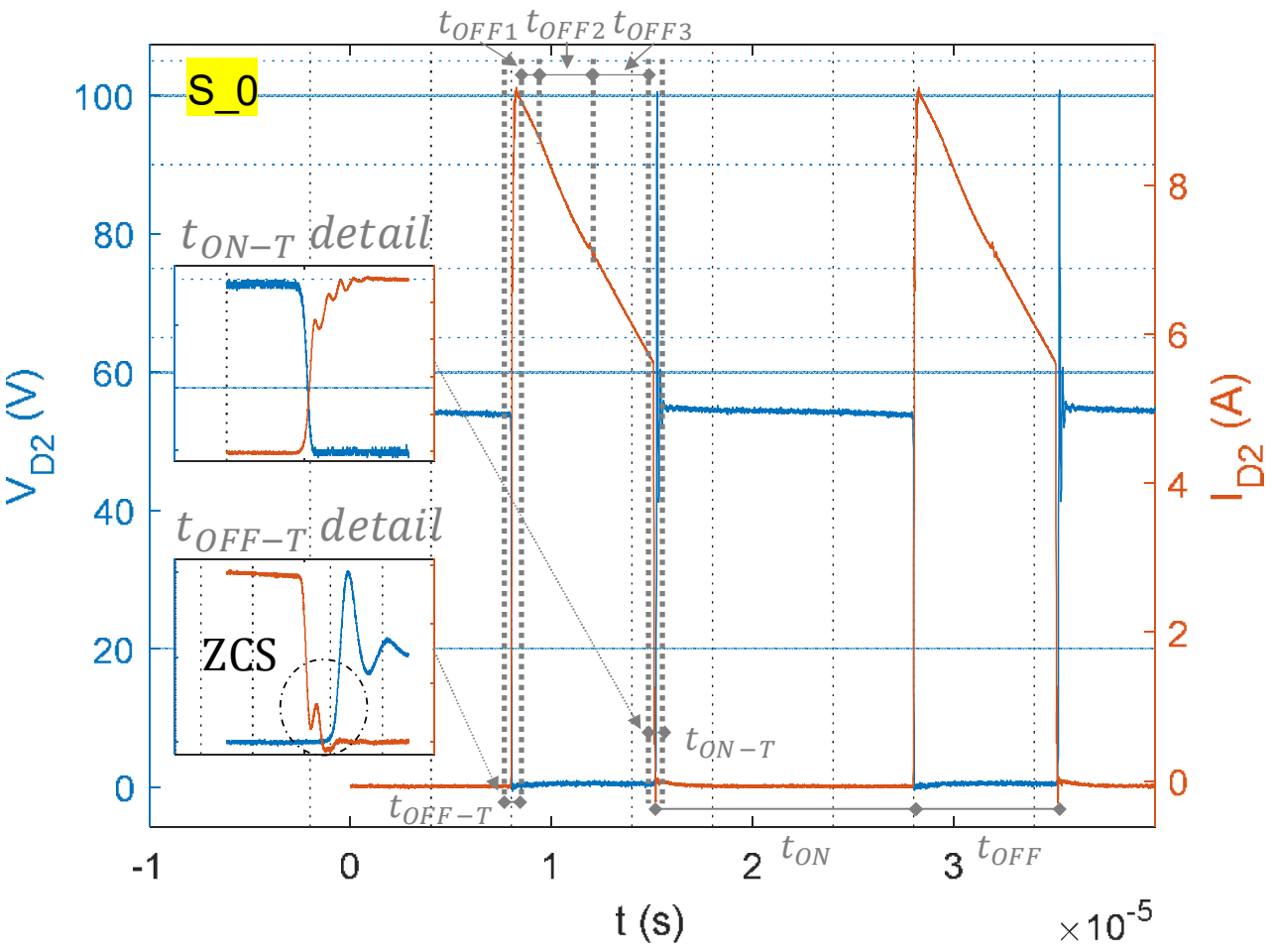

Figure 33. D2 waveforms being $V_i = 29.3\ V, V_o = 33.3\ V$ and $I_{string} = 6.75\ A$. VDS in blue color and IS in orange color

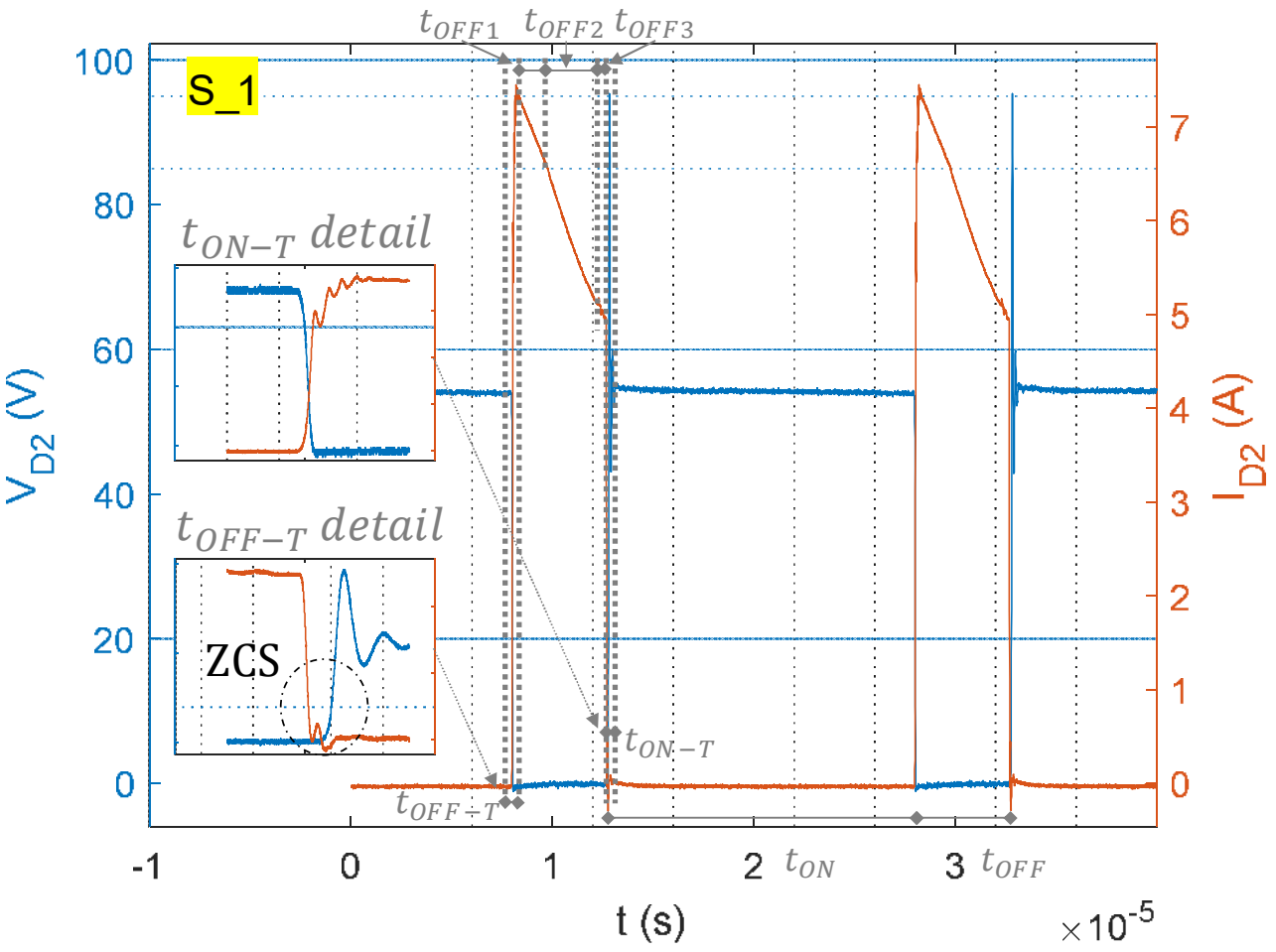

Figure 34. D2 waveforms being $V_i = 29.3\ V, V_o = 40.4\ V$ and $I_{string} = 5.569\ A$. VDS in blue color and IS in orange color

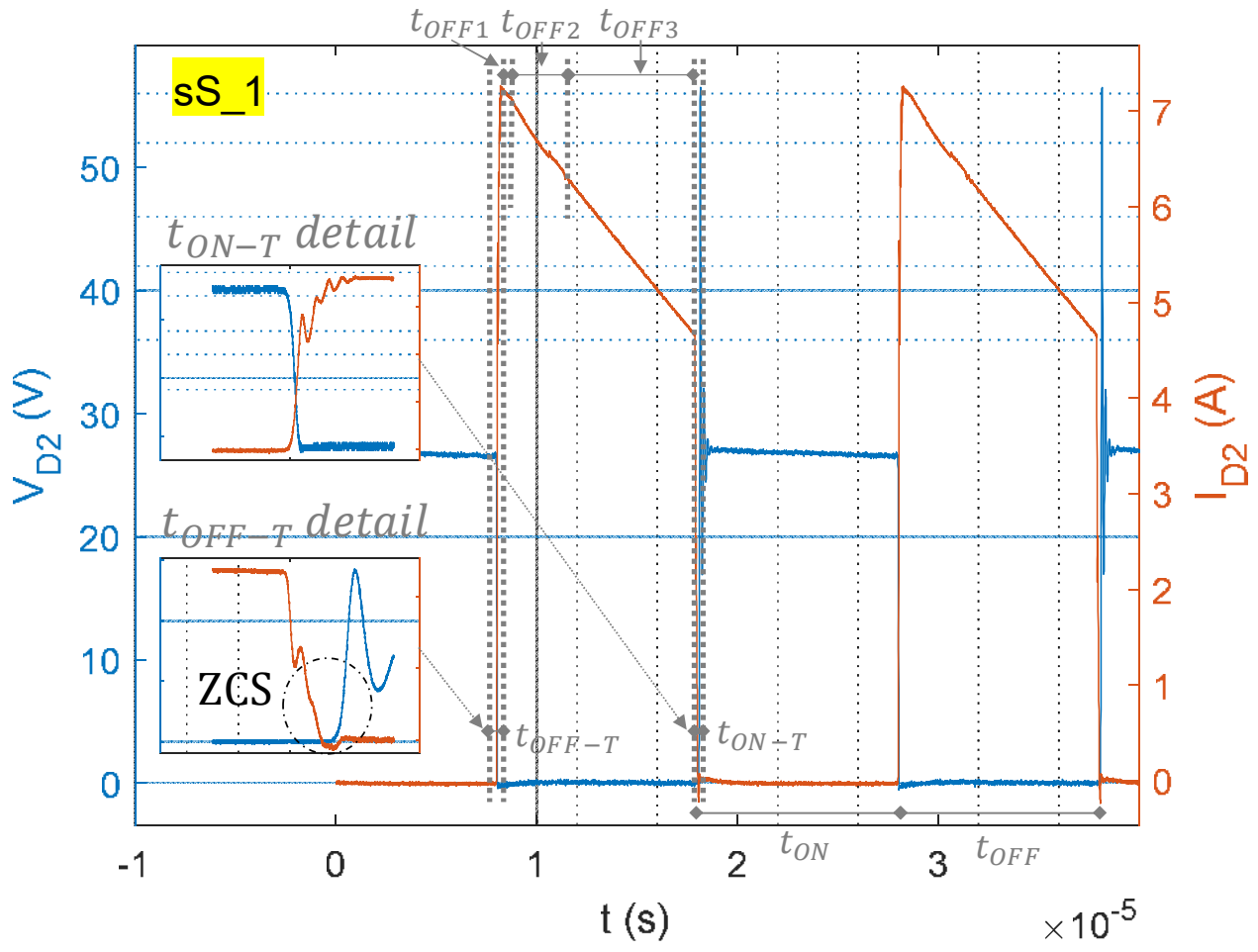

Figure 35. D2 waveforms being $V_i = 15\ V, V_o = 12.12\ V$ and $I_{string} = 5.569\ A$. VDS in blue color and IS in orange color

By comparing the behavior of the converter under the different scenario conditions, it can be concluded that the higher voltages in the S MOSFET and $D_1$ diode are obtained for the S_1 scenario when the output voltage of the converter is the highest. Regarding the current waveforms, due to the scenario S_0 has the highest string current, it is also reflected in the current waveforms of the components measured.

With the aim of validating the DMPPT concept, two converters are placed in cascade, connecting their outputs in series, as shown in Figure 36.

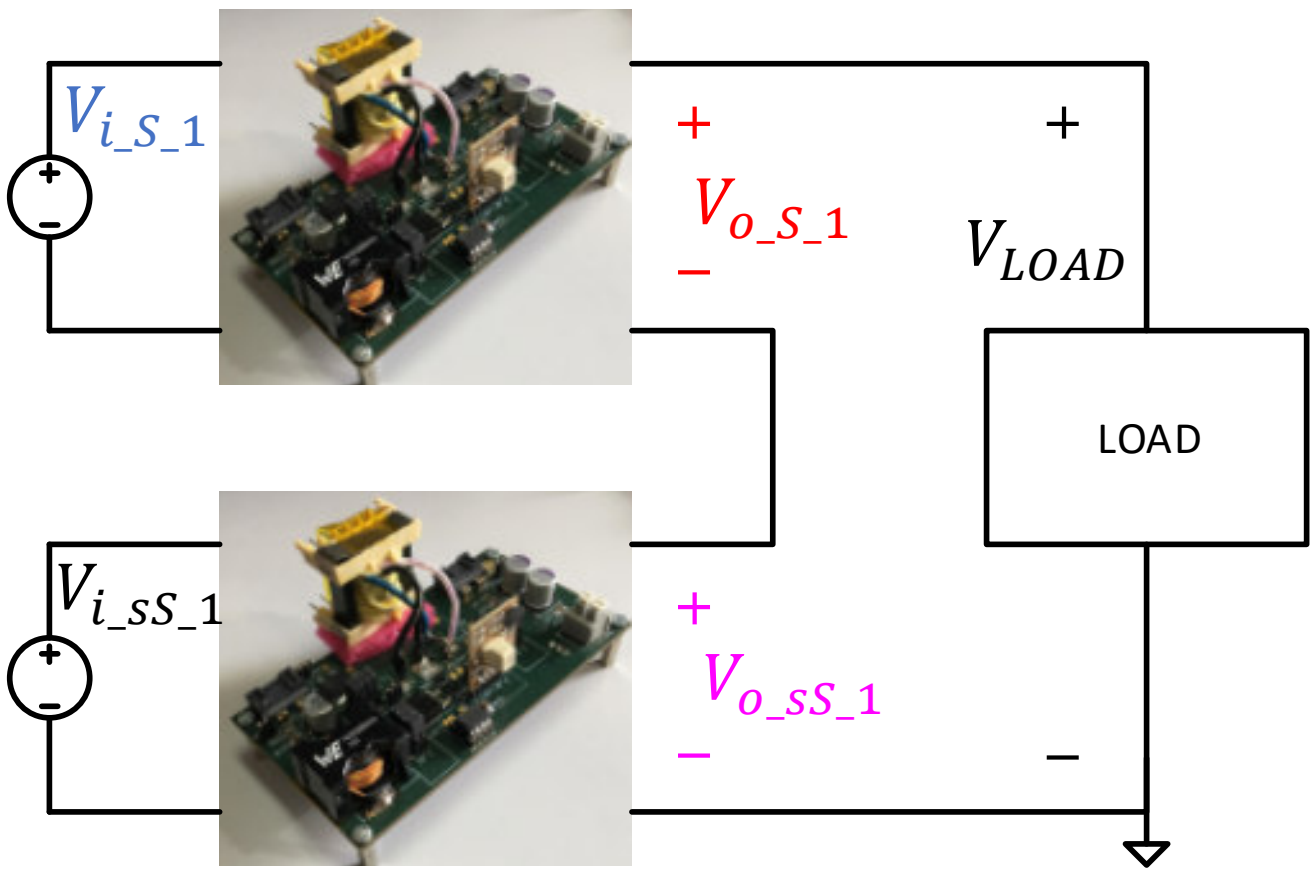

Figure 36. Setup scheme for the measurement connecting two AFZ converters in series

The measurement results are shown in Figure 37.

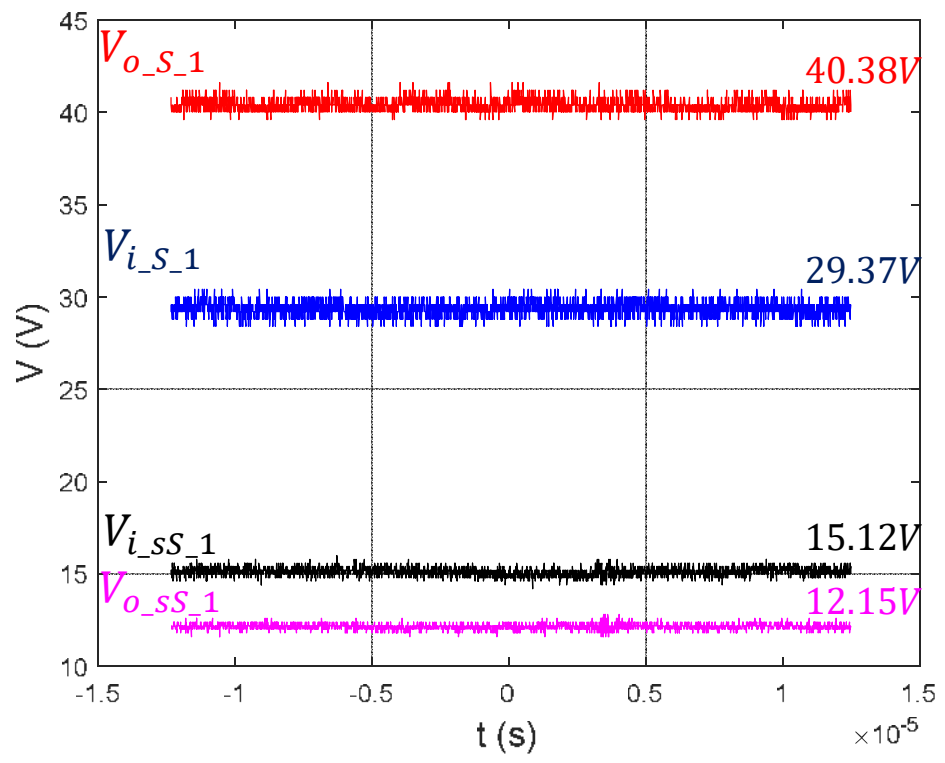

Figure 37. Measurement results connecting two AFZ converters in series. Input and output voltages of a converter connected to a non-shaded PV panel, $V_{i_S_1}$ and $V_{o_S_1}$; input and output voltages of a converter attached to a shaded PV panel, $V_{i_sS_1}$ and $V_{o_sS_1}$

Figure 37 shows the input and output voltages of two AFZ converters in series. The operating conditions are the same as for Scenario 1, where one converter emulates the behavior while it is connected to a shaded PV panel (sS_1), and the other one is connected to the same string, but to a non-shaded PV panel (S_1). It can be seen the correct operation of the AFZ converter while it is connected in series with another AFZ converter.

The measured waveforms correspond with the theoretical analysis exposed in Section II.B; therefore, those analyses are verified.

*2. Efficiency measurements*

The efficiency is measured using the wattmeter Yokogawa WT3000. Figure 38 shows the efficiency plots for the three voltage ratios defined in Section III.A.1, see Table VII. One for Scenario 0 (S_0) and two for Scenario 1 (sS_1 and S_1). Measurements have been carried out in the laboratory. The input power source of the converter is a DC power source, whereas an electronic load in current-control mode has been used as the load of the converter. The input and output voltage of the converter is kept constant during each power sweep. This behavior emulates different irradiation levels.

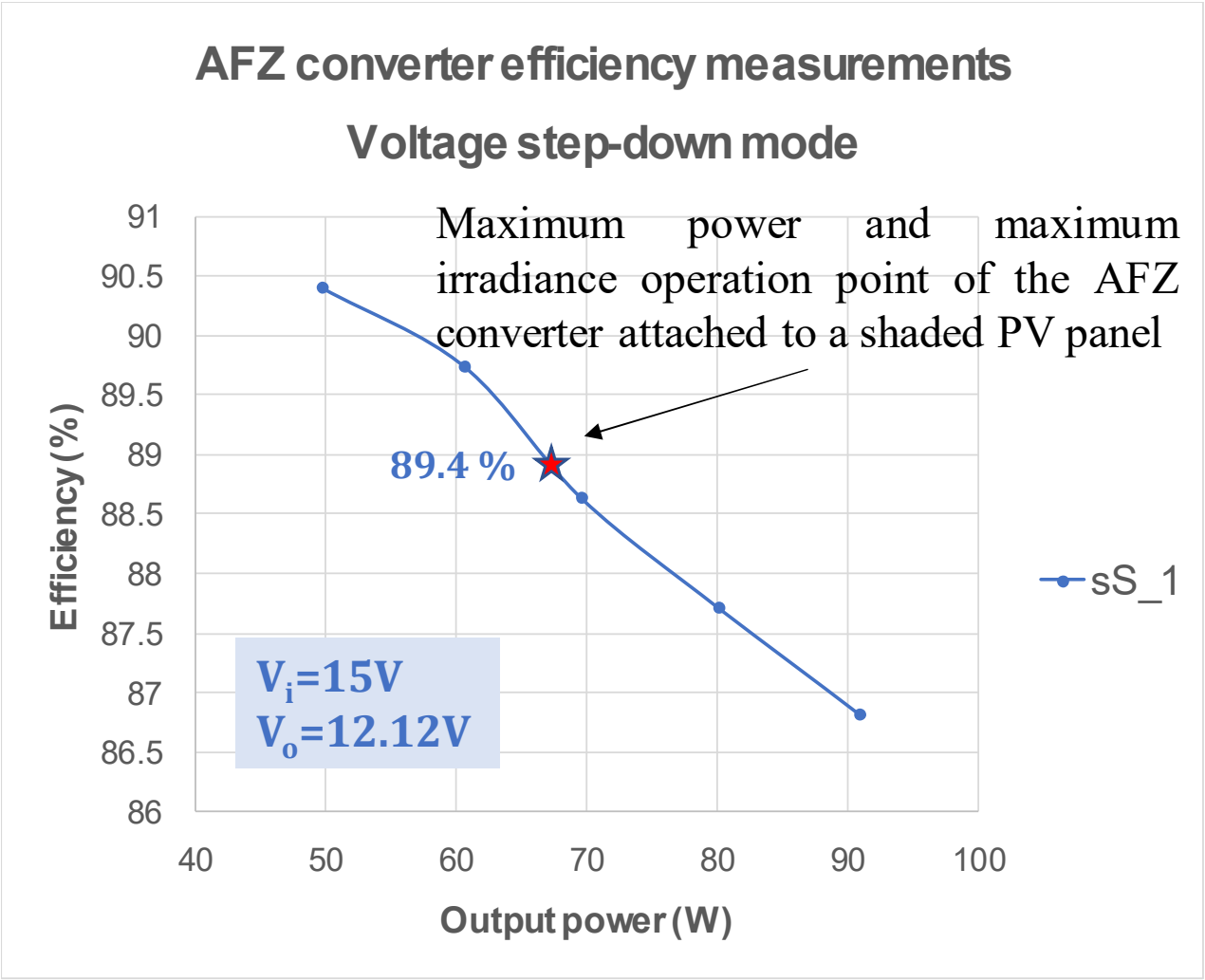

a)

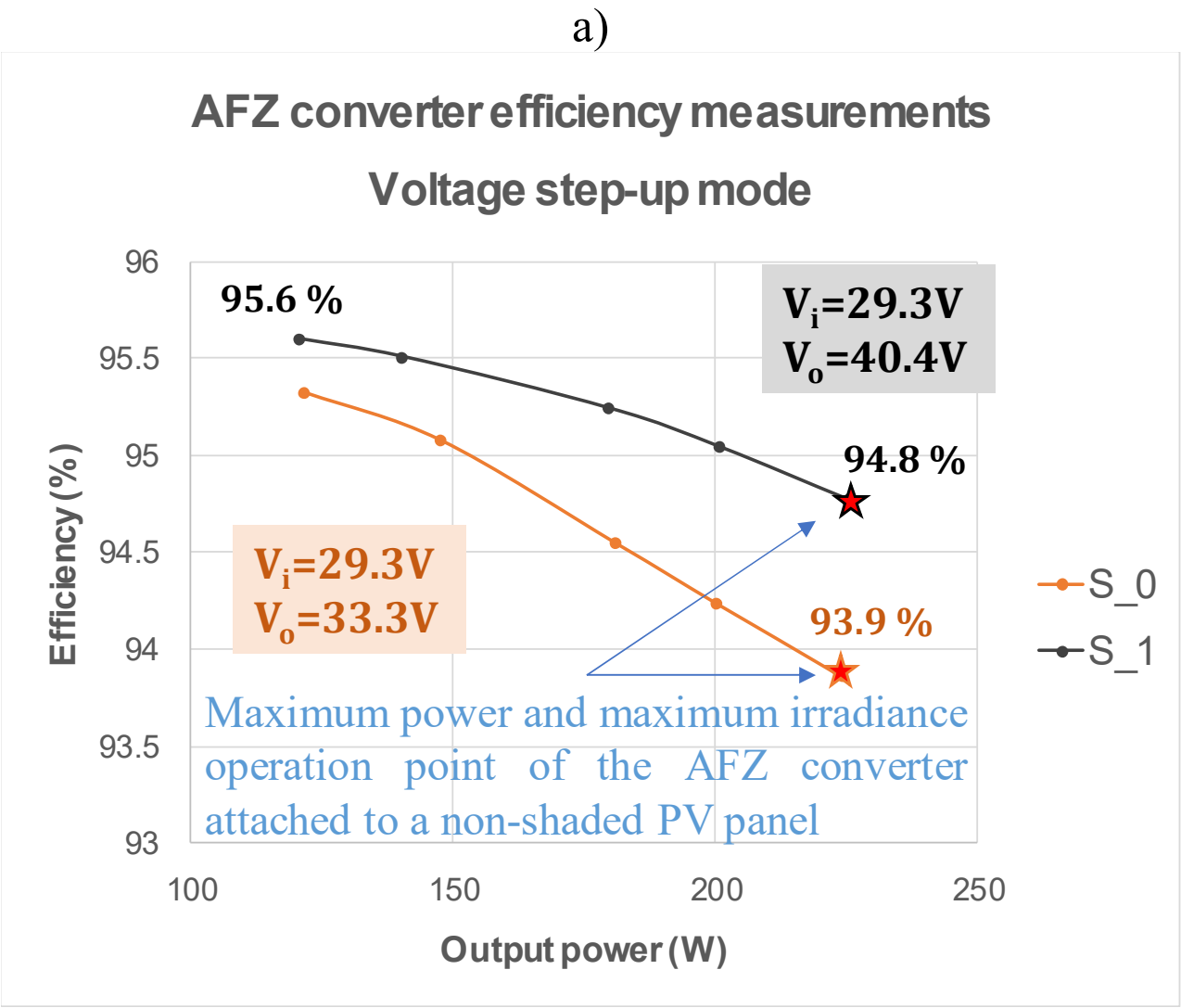

b)

Figure 38. AFZ efficiency measurements. a) Efficiencies for the converter attached to a shaded PV panel (sS_1); b) Efficiencies for the converter attached to a non-shaded PV panel (S_0, S_1)

Figure 38 a) depicts the power sweep carried out to the AFZ converter, for a fixed input and output voltages of 15 V and 12.12 V respectively, which correspond to the step-down mode voltages, see the Shaded PV panel voltages in Table VII. The efficiency of the AFZ converter on the sS_1 operation point is 89.4 %. This value is highlighted with a star. The other points of the power sweep curve are useful for analyzing the AFZ converter performance under other output power conditions and the same input and output voltages. Figure 38 b) shows the power sweep corresponding to the converter attached to a non-shaded PV panel when the AFZ converter operates on its voltage step-up mode. The input voltage for both scenarios is 29.3 V, and the output voltages are 33 V for S_0 and 40.4 V for S_1. The AFZ converter´s efficiencies on the operation point described in the case of study are 93.9 % for the S_0 scenario and 94.8 % for the S_1 scenario. Looking at the efficiency curves in Figure 38 b), the maximum measured efficiency is 95.6 %. This value is obtained with the higher voltage step-up, S_1 input and output voltages, and when the AFZ converter is managing around 120 W. As it can be seen, the higher the output voltage, the higher the efficiency of the AFZ converter. This fact reveals that the conduction losses, which are directly related to the output current, have more influence than the switching losses in the AFZ converter efficiency. The fact that the efficiency decreases as the output power increases reveals that the maximum efficiency point of the converter is at a lower output power than the maximum output power specified in both voltage step-up and voltage step-down operation modes. This type of tendency in the efficiency curve can be observed in papers such as [41], [42], [43], and [44].

Figure 39 depicts the power losses distribution ratios for 225 W output power and the AFZ converter under the S_0 conditions.

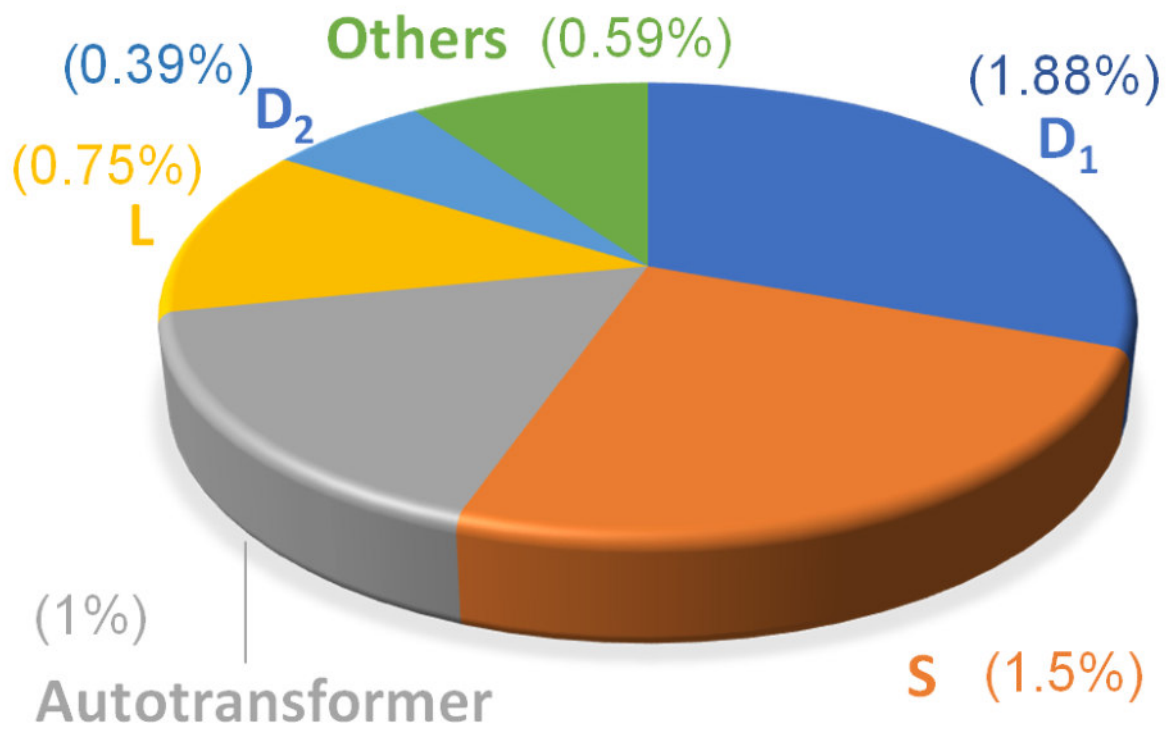

Figure 39. Power losses distribution ratios for the AFZ converter under S_0 conditions. The percentage represents the losses ratios for 225 W output power

The efficiency measurements close the study of the AFZ converter. Moreover, for a better understanding of the converter applicability, Table XI includes a comparison between the most relevant non-isolated voltage step-up and step-down converters for PV applications and the AFZ converter.

It can be seen that, although the efficiency obtained with the AFZ converter is higher than efficiencies achieved with other topologies for this application [31], [45], [46] and [47], recent works have reported higher efficiencies with the Non-Inverting Buck-Boost converter, see [15] and [28], and with passive ripple canceling circuit (PRCC) Cúk converter [48].

Table XI. Comparison among the AFZ converter and other module integrated converters available in the literature

| Parameters | AFZ | Buck-Boost [31] | Transformerless Buck-Boost [47] | ZETA based Buck-Boost [45] | SEPIC/ZETA [46] | Cúk with coupled inductor [31] | PRCC Cúk [48] | Non-Inverting Buck-Boost [28] |
|---|---|---|---|---|---|---|---|---|
| Magnetic components | 2 | 1 | 3 | 3 | 2 | 1 | 3 | 1 |
| Active switches | 1 | 1 | 1 | 1 | 2 | 1 | 1 | 4 |
| Drivers | 1 | 1 | 1 | 1 | 2 | 1 | 1 | 4 |
| Diodes | 3 | 1 | 3 | 2 | 2 | 1 | 1 | 0 |
| Dynamic performances | No-RHP zero | RHP zero | RHP zero | RHP zero | RHP zero | RHP zero | RHP zero | RHP zero |
| Efficiency (%) | 95.6 | 87.2 | 95.2 | 95.5 | 94.7 | 93.1 | 98 | 98.5 |

Not all converters shown in this table are designed for the same input voltage, output voltage and power range.

In comparison to the Non-Inverting Buck-Boost converter, the AFZ converter has the advantage of only including one active switch with only one driver. This fact simplifies the converter and improves the reliability of the overall system. Besides, the size of the output filter capacitors in the AFZ converter is smaller due to its output filter inductor.

Regarding PRCC Cúk converter, AFZ converter has non inverting output voltage and a higher number of magnetic components. Although the efficiency is higher than the AFZ converter, the power range and output current used for the design are more favorable.

It is worth noting that all the competitors suffer from RHP zero in their small-signal transfer functions, limiting their dynamic performance.

## IV. Conclusions

The analysis, design, and implementation of the Autotransformer Forward with type-Zeta resonant reset (AFZ) converter are presented in this paper. The theoretical analysis includes the steady-state and the small-signal study in continuous conduction mode.

The paper also includes a design procedure used to build a 225 W AFZ converter prototype for a specific field of application, consisting of a 100 kW grid-tied PV installation, with different mismatching ratios.

The experimental results verify the time domain theoretical analysis, whereas the frequency domain analysis is validated through simulation.

Experimental and simulation verification demonstrates the main characteristics of the AFZ converter that can be summarized as:

i) The type-Zeta resonant reset network allows the autotransformer core to be excited symmetrically in the first and third quadrants of the B-H plane, avoiding a third winding in the autotransformer. Therefore, the leakage inductance, the size, and the power losses are reduced. The resonant reset helps to avoid the use of additional snubber networks, reducing the number of components and increasing the efficiency of the converter.

ii) The autotransformer connection allows that only a part of the output power to be magnetically processed, as well as soft-switching features in the main diodes, $D_1$ and $D_2$, reaching ZVS during the $D_1$ turn-on transition and ZCS during the $D_2$ turn-off transition.

iii) The maximum measured efficiency is 95.6% (120 W output power and S_1 input and output voltages), which is in the range of the efficiencies obtained by the main competitors. The maximum efficiency achieved at 225 W, under the case of study conditions, is 94.8 %. This value corresponds to a converter attached to a non-shaded PV panel within a string with shaded PV panels.

iv) The voltage step-up and step-down capability of the AFZ converter gives the designer of a PV installation higher flexibility when selecting the optimum number of PV panels per string and the number of strings configuration.

v) A good dynamic performance, with no right-half-zero in their main small-signal transfer functions

vi) The low number of components, with only one active-switch and one driver, simplifies the converter, reduces the cost, and increases the reliability of the converter.

On the other hand, as the main disadvantages, the voltage stresses in the $D_1$ diode and MOSFET are high, and then devices with worse performances have to be selected. Also, although it is not a requirement for the field of application, the output of the converter is not isolated from the input, due to the use of an autotransformer.

ACKNOWLEDGMENT

This work has been partially supported by the Spanish Ministry of Economy and Competitiveness and FEDER funds through the research project "Storage and Energy Management for Hybrid Electric Vehicles based on Fuel Cell, Battery and Supercapacitors" - ELECTRICAR-AG - (DPI2014-53685-C2-1-R), and also through the research project CONEXPOT (DPI2017-84572-C2-2-R) and EPIIOT (DPI2017-88062-R).